\documentclass[epj]{svjour}
\usepackage{graphicx}
\usepackage{epsfig}
\usepackage{amsmath,amssymb}
\usepackage[percent]{overpic}

\begin{document}

\title{Synchronisation in networks of delay-coupled type-I excitable
  systems}
\author{Andrew Keane\inst{1} \and Thomas Dahms\inst{1} \and Judith Lehnert\inst{1} \and Sachin Aralasurali Suryanarayana\inst{1,2} \and Philipp H{\"o}vel\inst{1,3,4} \and Eckehard Sch{\"o}ll\inst{1} } 
\institute{Institut f{\"u}r Theoretische Physik, Technische Universit{\"a}t Berlin, 10623 Berlin, Germany \and
  Department of Physics, Indian Institute of Technology Bombay, Mumbai, India \and
  Bernstein Center for Computational Neuroscience, Humboldt-Universit{\"a}t zu Berlin, Philippstra{\ss}e 13, 10115 Berlin, Germany \and
  Center for Complex Network Research, Northeastern University, 110 Forsyth Street, Boston, MA 02115, USA
   }
\mail {schoell@physik.tu-berlin.de}

 
\date{\today}

\abstract{
  We use a generic model for type-I excitability (known as the SNIPER
  or SNIC model) to describe the local dynamics of nodes within a
  network in the presence of non-zero coupling delays. 
  Utilising the method of the Master
  Stability Function, we investigate the stability of the zero-lag
  synchronised dynamics of the network nodes and its dependence on the
  two coupling parameters, namely the coupling strength and delay
  time. Unlike in the FitzHugh-Nagumo model (a model for type-II excitability), 
  there are parameter ranges where the stability of synchronisation depends on the
  coupling strength and delay time.
  One important implication of these results is that there exist 
  complex networks for which the adding of inhibitory links in a small-world fashion may   not only lead to a loss of stable synchronisation, but may also
  restabilise synchronisation or introduce multiple transitions
  between synchronisation and desynchronisation. To underline the scope of our
  results, we show using the Stuart-Landau model that such multiple
  transitions do not only occur in excitable systems, but also in
  oscillatory ones.
}

\maketitle

\section{Introduction}
\label{section:intro}


Studies on complex networks have gained much attention in recent times
from various fields of research \cite{ALB02a,NEW03,BOC06a}. In
particular, the dynamics on networks of nonlinear elements (or
\textit{nodes}) has become a very active area
\cite{TIM02a,STE06a,ASH07,FLU10b,LEH11,ENG11,KAN11,OME11,DAH12,HAG12,NEP12}. 
Here, we consider the dynamics of neurons as a network of nonlinear
excitable nodes. While the model we consider is generic for
excitability, the local dynamics being modelled in this context is the
``firing'' mechanism of neurons (i.e. the rise and fall of the
electrical membrane potential). This mechanism is the basis for neural
coding and information transfer or cell-to-cell
communication~\cite{PER68}. The type of network we primarily focus on
in our investigations is the small-world (SW) network~\cite{WAT98},
which has a short average path distance between nodes, as well as a
large degree of clustering (in other words, many triangles in the
network structure). These properties are found in many kinds of
real-world structures, such as the collaboration between film actors,
power grids, the World Wide Web and social relationships
\cite{WAT98,ADA99,ALB02a,NEW03}. In particular, large-scale cortical
networks also show these properties \cite{SPO00,SPO06}. The brain has
an architecture enabling both efficient global and local communication
between neurons \cite{LAT01}, which is captured well by the SW model.


The interest in the synchronisation of neurons emerges from the role
it plays in information processing and perception, as well as in
epilepsy and similar diseases \cite{TRA82,ROE97,ENG01a,PIK01,MEL07}. 
While there are several types of synchronisation, the focus
here will be on \textit{zero-lag} synchronisation. This means that the
dynamics of all nodes in the network are not only identical but also
temporally in-phase.


Inhibition plays an important role in the nervous system \cite{HAI06}.
Here, when constructing a network we begin with a regular ring network
of excitatory links and, as in Ref.~\cite{LEH11}, we add long-range
inhibitory links into the network structure. This creates a SW network
of the form proposed in Ref.~\cite{NEW99b}. The introduction of inhibitory
links and an increasing of their density in the network was shown in
Ref.~\cite{LEH11} to result in a loss of stable synchronisation.

In this paper, we consider delay-coupled networks of type-I
excitability, in contrast to Ref.~\cite{LEH11} where type-II 
was considered using FitzHugh-Nagumo local dynamics, 
and find qualitative differences for a particular range of
coupling parameters. The most interesting implication of this is that
for some network topologies with this type of excitability increasing
the density of inhibitory links may not just lead to one transition
from stable to unstable synchronisation, but may lead to multiple
transitions. This means that increasing the density of inhibitory
links in the network may stabilise synchronisation, rather than
destabilise it. 

In the following section, we introduce the model and the network
topologies considered in the present study. In Sec.~\ref{section:MSF} we use the master
stability function to investigate the stability of arbitrary
synchronised networks with given coupling parameters (i.e. the
coupling strength and the length of delay between coupled nodes). The
master stability function is calculated in Sec.~\ref{section:small_tau} for networks coupled
within a range of small delay time and coupling strength, which
delivers the main results of the paper - namely, the existence of
synchronised states that have different stability conditions compared
to coupling with larger delay times. The implications these results
may have for specific complex networks are discussed in
Sec.~\ref{section:implications}. In Sec.~\ref{sec:stuart-landau}, we
study networks of coupled Stuart-Landau oscillators and demonstrate
that multiple transitions between synchronisation and
desynchronisation can also appear in networks of oscillatory nodes and
are not limited to excitable systems.

\section{Model}
\label{section:model}

In order to model excitability, the system must have a rest state,
which corresponds to a stable fixed point. Small perturbations from
the rest state allow for the creation of a large excursion in the
phase space, which corresponds to the system entering an excited
state before returning to the rest state. In the context of 
neurodynamics, this is the firing state of the neuron \cite{IZH00}.


Neurons can exhibit different excitability properties. In 1948,
Hodgkin classified two types of neural excitability \cite{HOD48}. Both
types can be ordered depending on the type of bifurcations that occurs
when the bifurcation parameter is changed such that the
system makes a transition from the excitable to the oscillatory
regime:

Type-I neurons can generate action potentials of arbitrarily low
frequency. This kind of behavior occurs near a saddle-node infinite
period (SNIPER) bifurcation, also known as the SNIC bifurcation
(saddle-node bifurcation on invariant cycle). The arbitrarily low
frequency coincides with the period of the limit cycle going to
infinity as the bifurcation parameter approaches a critical value,
where the bifurcation occurs.

Type-II neurons are associated with a supercritical Hopf bifurcation.
The frequencies of the action potentials lie within a certain non-zero
range, while the amplitude of the limit cycle approaches zero with the
bifurcation parameter.

As our model for type-I excitability we consider a generic normal-form of a
SNIPER bifurcation~\cite{HU93a,HIZ07}. This model is mathematically
represented by two first-order differential equations:
\begin{equation}
  \label{eq:model}
  \mathbf{f}(\mathbf{x})=\begin{pmatrix}
    \dot{x}\\
    \dot{y}\end{pmatrix}
  =\begin{pmatrix}
    x(1-x^2-y^2)+y(x-b)\\
    y(1-x^2-y^2)-x(x-b)\end{pmatrix}.
\end{equation}
In polar coordinates $(x=r \cos{\varphi}, y=r \sin{\varphi})$ Eq.~(\ref{eq:model}) reads
\begin{eqnarray}
  \label{eq:polarcoord}
\dot{r}&=&r\left( 1-r^2 \right)\\
\dot{\varphi}&=&b-r\cos\varphi.
\end{eqnarray}
$b>0$ is the bifurcation parameter. This bifurcation parameter influences the type of dynamics 
and determines where in the $(x,y)$-plane
the fixed points are located, as discussed below.


An important feature that must be taken into account in the
mathematical model is that the effect one neuron has on another is not
instantaneous. The signals sent between neurons have a finite
propagation speed and take a certain time to arrive at the destination
neuron, giving rise to a delay time $\tau$.

The dynamics of a network of $N$ elements (labelled $i=1,...,N$) is
written:
\begin{equation}
  \label{eq:neteqns}
  \dot{\mathbf{x}}_i = \mathbf{f}(\mathbf{x}_i) + \sigma \sum_{j=1}^N G_{ij} \mathbf{H} (\mathbf{x}_j(t-\tau)-\mathbf{x}_i(t)),
\end{equation}
where $\mathbf{f}(\mathbf{x}_i)$ is the local dynamics as described by
Eq.~(\ref{eq:model}) for each element $\mathbf{x}_i=(x_i,y_i)$.
$G_{ij}$ determines the matrix $\mathbf{G}$ for the network structure, showing which
elements are coupling together, and $\mathbf{H}$ is the coupling
function. $\mathbf{H}$ is taken to be
the $2\times 2$ identity matrix; this means that the $x$-variable at
time $t$ is coupled with the $x$-variable at time $t-\tau$, the
$y$-variable at time $t$ is coupled with the $y$-variable at time
$t-\tau$, but there is no cross-coupling between the $x$- and $y$-variable.
The coupling parameters, which are identical for all connections, are the coupling strength $\sigma$ and delay
time $\tau$.

For synchronous dynamics, the $2(N-1)$ constraints
$\mathbf{x}_{s}\equiv\mathbf{x}_{1}=\mathbf{x}_{2}=\cdots=\mathbf{x}_{N}$
define a 2-dimensional synchronisation manifold (SM) within the
$2N$-dimensional phase space and the coupled system is reduced to an
effective single system with feedback:
\begin{equation}
  \label{eq:sync_eqn}
  \dot{\mathbf{x}}_{s} = \mathbf{f}(\mathbf{x}_{s}) + \sigma\mathbf{H}(\mathbf{x}_{s}(t-\tau)-\mathbf{x}_{s}(t)),
\end{equation}
with unity row sum $\sum_j G_{ij}=1$, so that the nodes all receive
the same level of input while they are synchronised. Any non-unity but
constant row sum can be rescaled using the coupling strength $\sigma$.

With the bifurcation parameter $b<1$ there exists an unstable focus at
the origin, as well as a saddle point and a stable node situated on the
unit circle at $(x_i,y_i)=(b,\sqrt{1-b^{2}})$ and $(b,-\sqrt{1-b^{2}})$, 
respectively. At $b=1$ the saddle point and stable node
collide, so that for $b>1$ a limit cycle exists along the unit circle.
In the case of $b<1$ (excitable regime), a perturbation which pushes the system 
from the stable node beyond the saddle point can result in a single oscillations
along the unit circle. Delayed coupling can induce a homoclinic
bifurcation, such that a limit cycle is produced that bypasses the
saddle point and stable node \cite{HIZ07,AUS09}. Here, we will 
consider the excitable regime with $b=0.95$ and investigate for which topologies the synchronised dynamics of complex networks is stable.

In this paper, two kinds of complex networks are considered: SW
networks and Erd\H{o}s-R\'{e}nyi random networks \cite{ERD59}. We
construct the SW networks as a variation to the method proposed in
Ref.~\cite{WAT98}, introduced in Refs.~\cite{MON99,NEW99b}: (i) Each of the $N$
nodes in a one-dimensional ring is given excitatory links to its $k$
nearest neighbours on each side. Note that in terms of the matrix
$\mathbf{G}$, an excitatory link between the $i^{\textnormal{th}}$ and
$j^{\textnormal{th}}$ node means that $G_{ij}>0$, while for an
inhibitory link $G_{ij}<0$. (ii) For each of the $kN$ links we add
with a probability of $p$ another inhibitory link with strength -1
between two randomly chosen nodes (i.e. on average $pkN$ randomly distributed
inhibitory links). (iii) Self-coupling and multiple links between the
same two nodes are not allowed. (iv) Finally, the entries in each
row of $\mathbf{G}$ are normalised to ensure a unity row sum. If a row
sum is equal to zero, then the network realisation is discarded. 

The random networks are constructed with $kN$
excitatory links between randomly chosen nodes,
as well as $pkN$ randomly distributed inhibitory links, so that these
may be compared to a SW network with the same parameters $p$, $k$ and
$N$. Because the construction of both kinds of complex networks involves random processes,
there will be many possible realisations of either kind of network
with a given set of parameters.

In the following, we determine stability of the synchronised dynamics
and compare the stability of these different kinds of networks.

\section{Stability of Synchronisation}
\label{section:MSF}

The Master Stability Function (MSF) is a tool that represents a measure of
the stability of a synchronised state for given delay-coupling
parameters in terms of the topology of an arbitrary network
\cite{PEC98}. The master stability equation, given by
\begin{equation}
  \label{eq:mse}
  \delta\dot{{\mathbf{x}}}(t) = [D\mathbf{f}(\mathbf{x}_{s}) - \sigma\mathbf{H}]\delta{\mathbf{x}}(t)
  + (\alpha+i\beta)\mathbf{H}\delta{\mathbf{x}}(t-\tau),
\end{equation}
is found by linearising Eq.~(\ref{eq:neteqns}) around
Eq.~(\ref{eq:sync_eqn}) and can be used to calculate the largest
Lyapunov exponent $\Lambda(\alpha,\beta,\sigma,\tau)$,
called the MSF.
$\delta{\mathbf{x}}$ is the perturbation of $\mathbf{x}$ away from the
SM (i.e. $\mathbf{x}=\mathbf{x}_s+\delta\mathbf{x}$) and
$D\mathbf{f}(\mathbf{x}_s)$ is the Jacobian matrix of
Eq.~(\ref{eq:model}) evaluated on the SM. Important for this approach is that,
whereas one can calculate Lyapunov exponents for a specific network
topology using the eigenvalues of the matrix $\mathbf{G}$, one
considers here a continuous complex parameter $\alpha+i\beta$
representing the complex plane of possible eigenvalues scaled by the
coupling strength $\sigma$ (i.e. $\alpha+i\beta$ is a continuous
parametrisation of $\sigma\nu_j$, where $\nu_j$ are the eigenvalues of
$\mathbf{G}, j=1,...,N$). Thus, one can calculate the Lyapunov
exponents for a region of the $(\alpha,\beta)$-plane which gives
sub-regions of stability, where $\Lambda<0$, and instability, where
$\Lambda>0$. It is then easy to compare the synchronous stability of
various networks by simply observing whether any of their eigenvalues
fall inside an unstable region of the $(\alpha,\beta)$-plane. If all
the eigenvalues lie within stable regions, then perturbations away
from the SM will decay exponentially.

Because of the unity row sum condition, $\mathbf{G}$ always has an
eigenvalue $\nu_1=1$. This longitudinal eigenvalue (all others are
called transversal) corresponds to perturbations within the SM and
$\Lambda(\sigma\nu_1,0,\sigma,\tau)$ is always zero because we are
looking at periodic dynamics. As such, it is only the transversal
eigenvalues that are important for determining the stability of
synchronisation.

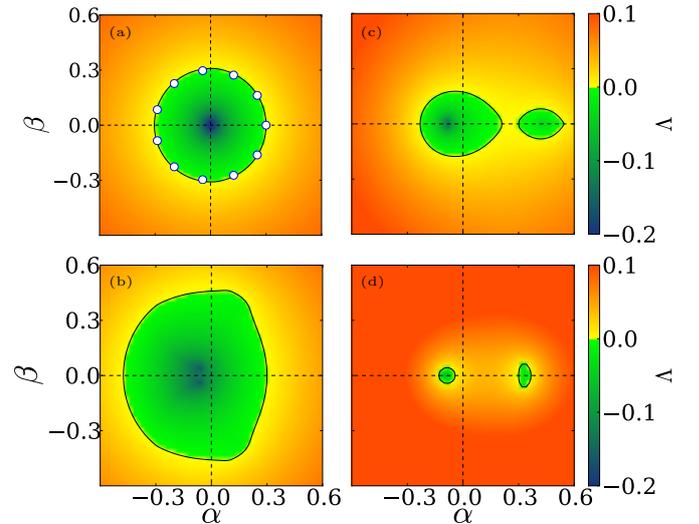
\begin{figure}
  \hspace{0.1mm}
  \begin{overpic}[width=0.99\columnwidth,height=0.82\columnwidth]{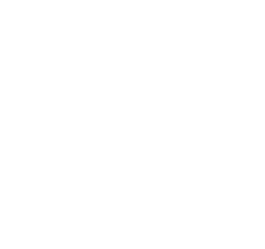}
  \put(1.5,45){\includegraphics[scale=0.212]{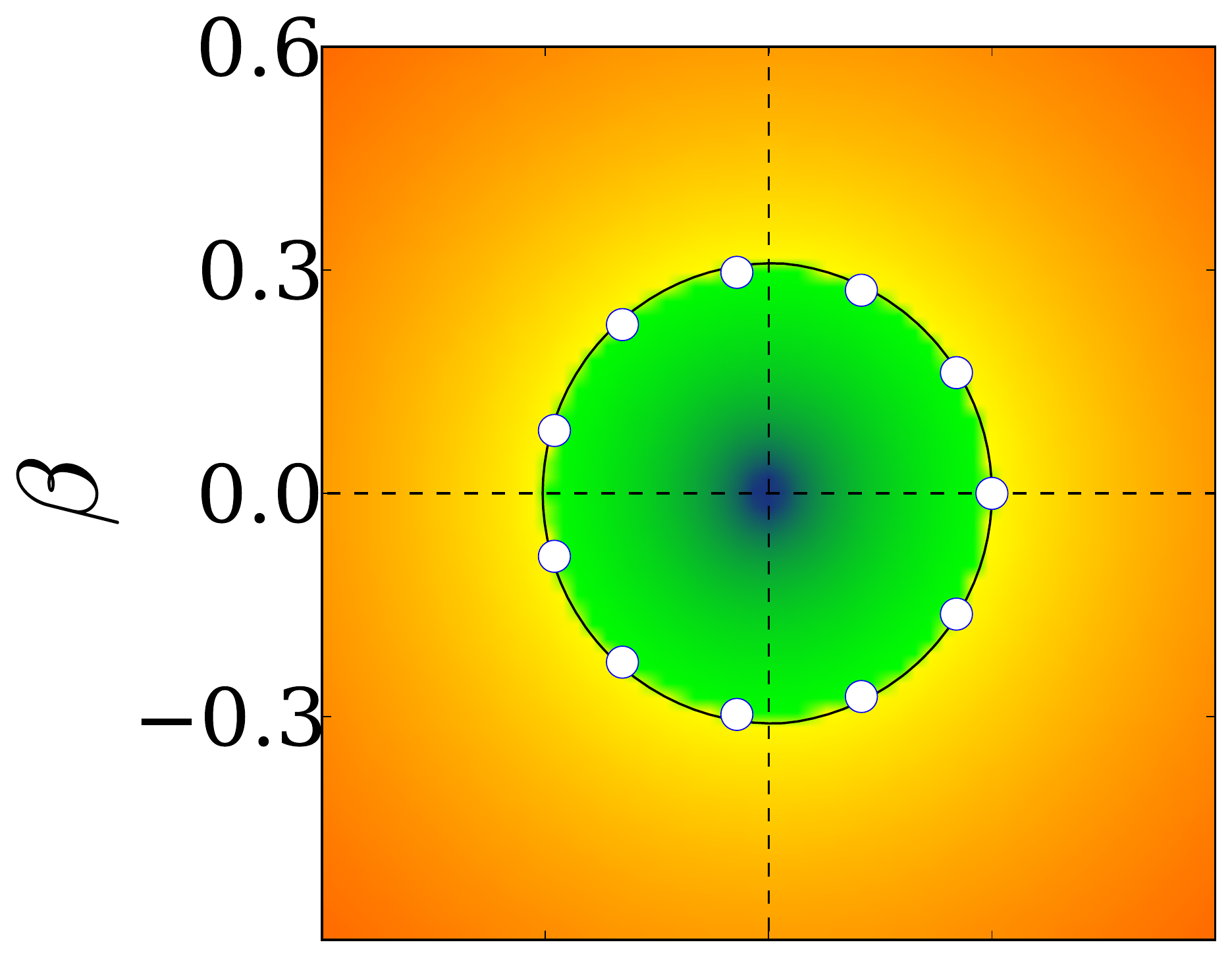}}
  \put(15,76){\tiny(a)}
  \put(0,1){\includegraphics[scale=0.24]{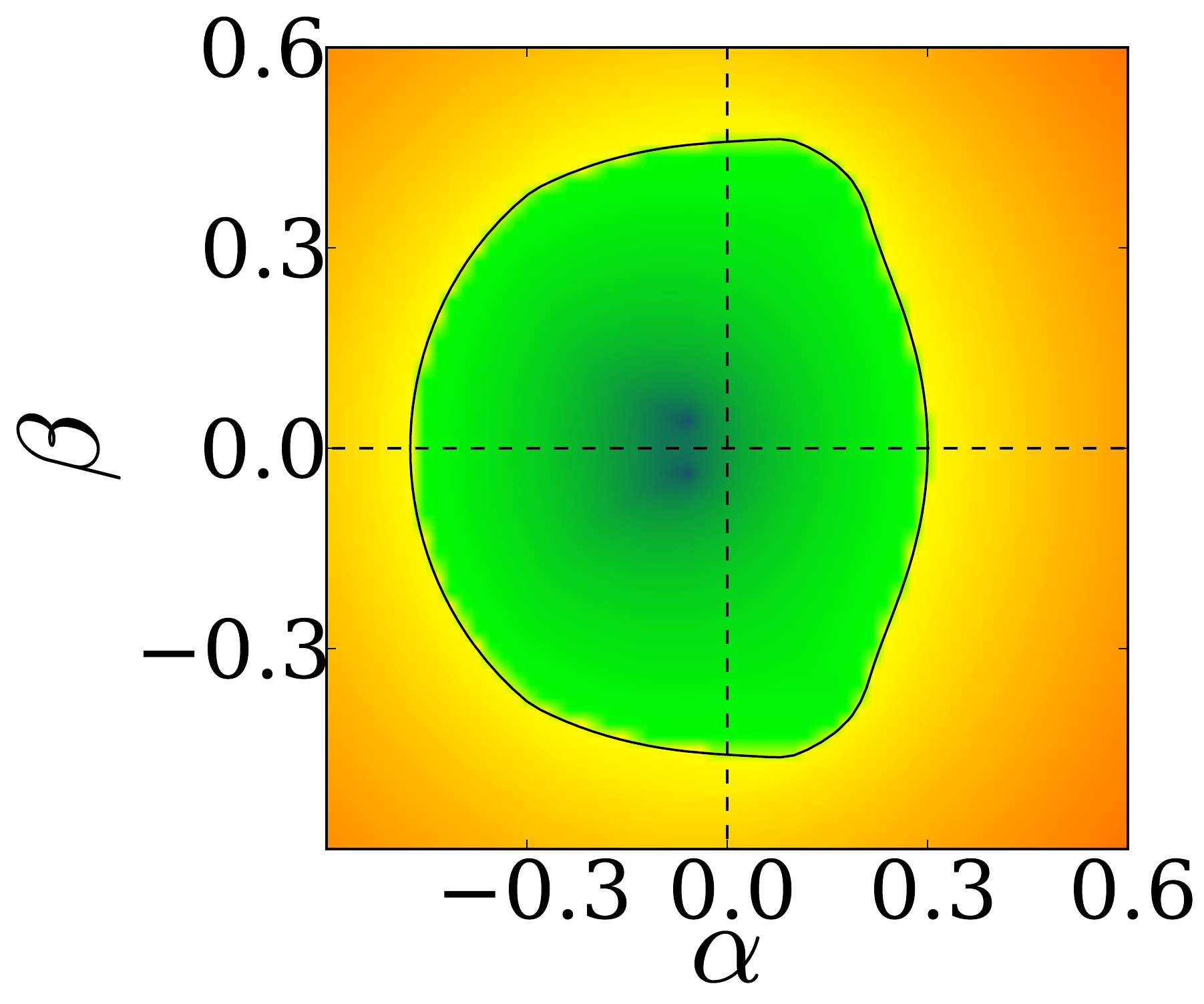}}
  \put(15,38){\tiny(b)}
  \put(51,43.8){\includegraphics[scale=0.24]{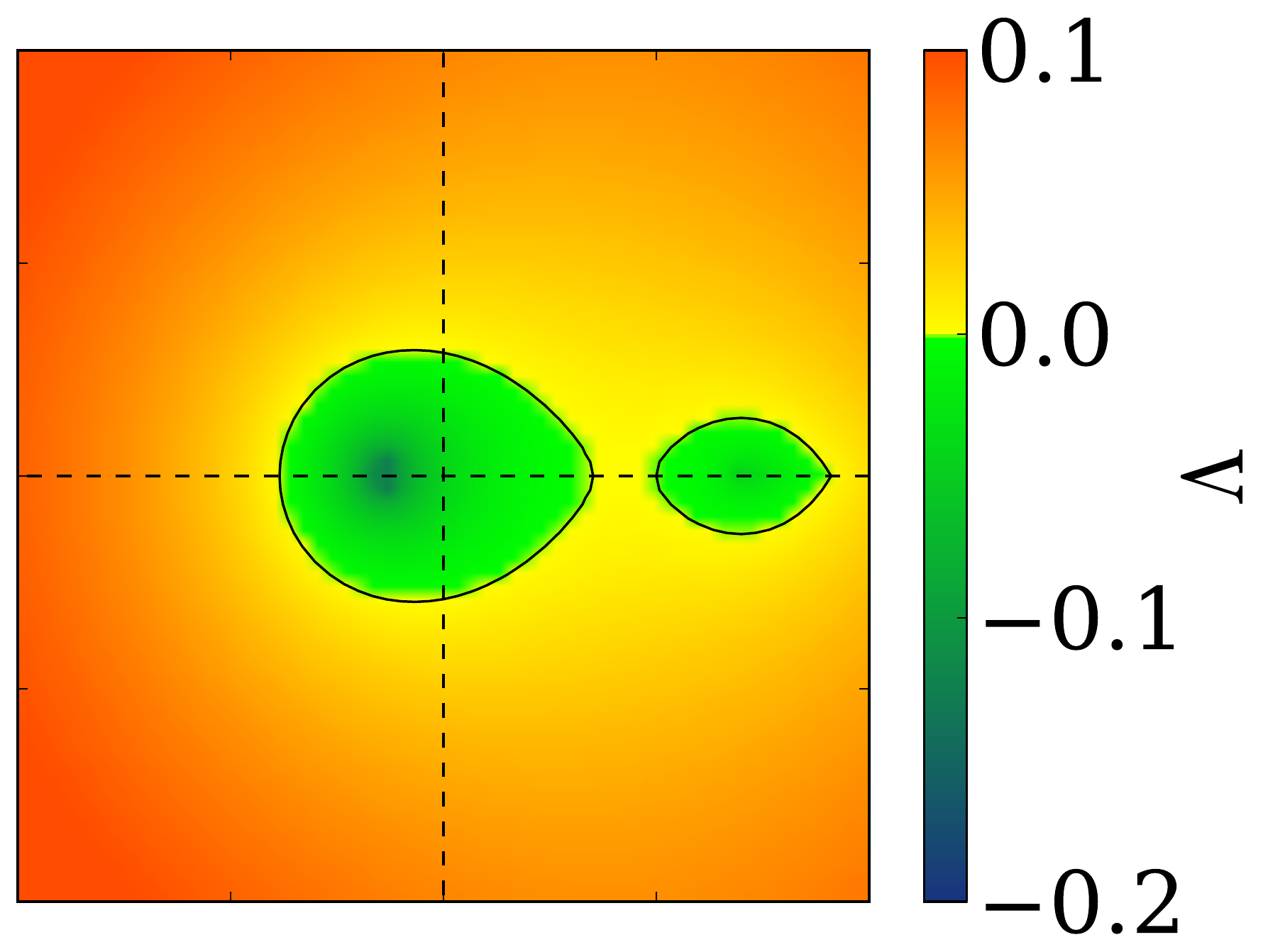}}
  \put(53,76){\tiny(c)}
  \put(51,1){\includegraphics[scale=0.24]{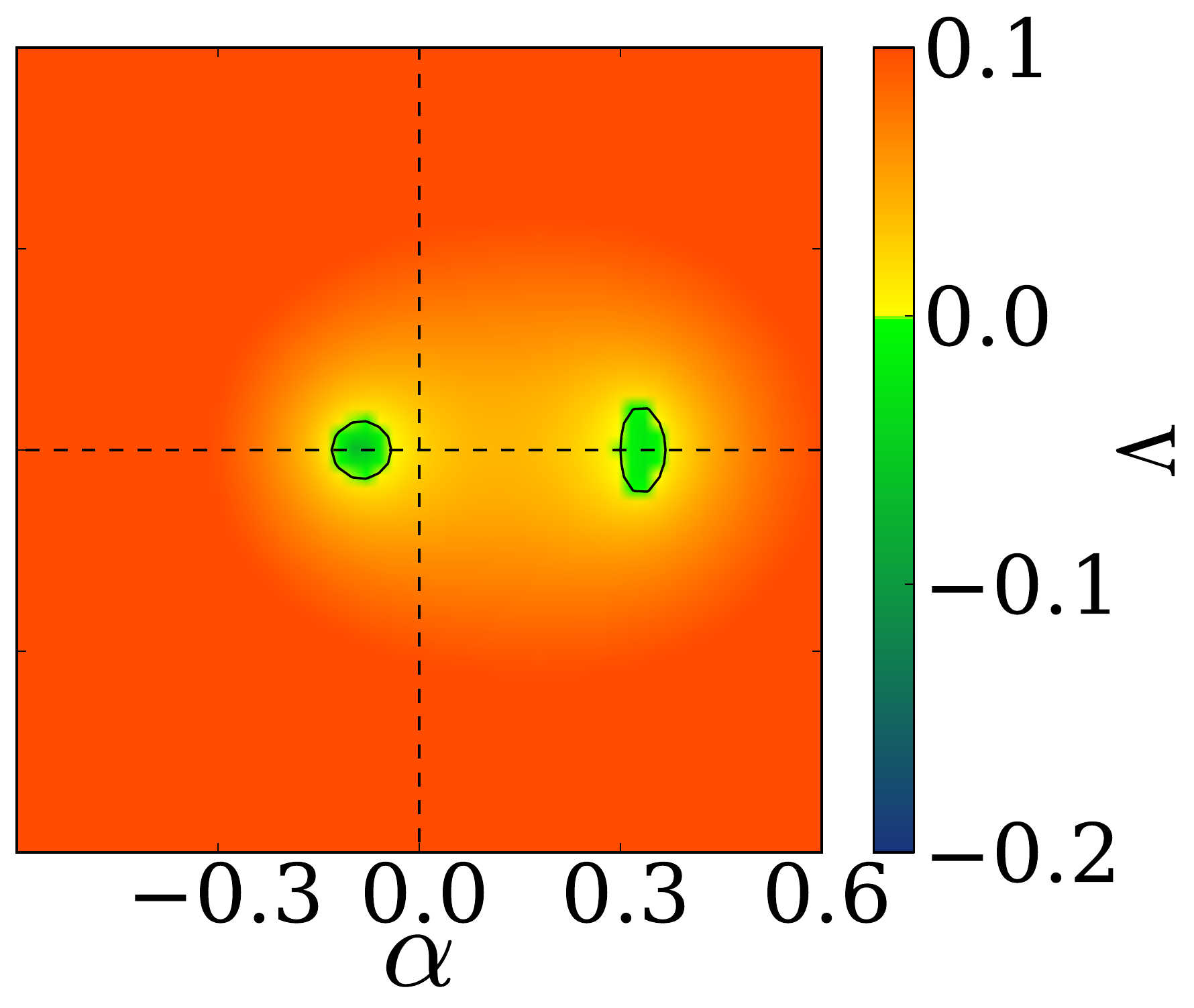}}
  \put(53,38){\tiny(d)}
  \end{overpic}
  \caption{(Colour online) Master stability function $\Lambda$ for coupling parameters $\sigma=0.3$ and (a)
    $\tau=10$, (b) $\tau=7$, (c) $\tau=6.5$, and (d) $\tau=6$.
    $\alpha$ and $\beta$ are the real and imaginary parts of the scaled eigenvalues of the coupling matrix,
    $\sigma\nu_k$, respectively, and $\Lambda$ is the largest Lyapunov
    exponent. White dots in panel (a) represent the scaled eigenvalues
    of a unidirectional ring of 11 nodes ($\sigma\nu_{1},\ldots,\sigma\nu_{11}$).
    $b=0.95$.}
  \label{fig:msf_symmetry_breakdown}
\end{figure}

Figure~\ref{fig:msf_symmetry_breakdown}(a) shows the MSF of the SNIPER model
with coupling parameters $\sigma=0.3$ and $\tau=10$. The white dots
represent the eigenvalues of the matrix $\mathbf{G}$ for a
unidirectional ring of 11 nodes. All eigenvalues are within the stable
region of the MSF, thus the synchronisation of all 11 nodes coupled
with these parameters is stable.

According to Ref.~\cite{FLU10b}, for $\tau$ in the order of the
system's characteristic time scale (in this case, the period of the
oscillations) and above, the MSF will always tend towards a rotational
symmetry. Examples such as the one above in
Fig.~\ref{fig:msf_symmetry_breakdown}(a) confirm these general findings for the SNIPER
model. When calculating MSFs for a large fixed $\tau$ while varying $\sigma$
it becomes evident that the size of the stable region is scaled by
$\sigma$, so that the stable region can be estimated very well by the
circle $S((0,0),\sigma)$ (that is, a circle centred at the origin with
a radius $\sigma$). Large values of $\tau$, however, despite
influencing the Lyapunov exponents quantitatively, do not affect the
shape of the stable region.

This rotational symmetry of the MSF was also found for type-II neurons,
modelled as FitzHugh-Nagumo oscillators \cite{LEH11}. In this case
$\sigma$ and $\tau$ do not affect whether the eigenvalues fit into the
stable region of the MSF, so that only the topology of a network is
important for the stability of its synchronisation. Furthermore,
because Gershgorin's circle theorem \cite{GER31} guarantees that the
eigenvalues of a network's coupling matrix with no self-coupling and
purely excitatory coupling (i.e. $G_{ii}=0$ and $G_{ij}\geq 0, 1\leq
i,j\leq N$) lie within the unit circle on the complex plane, the
synchronisation of such a network will always be stable. Finally,
if additional inhibitory links are introduced with probability $p$
to construct a small-world network as described in Sec.~\ref{section:model},
phase transitions from stable to unstable synchronisation are found 
with increasing probability of inhibition $p$ \cite{LEH11}.
All these results also apply for the SNIPER neurons with sufficiently
large $\tau$.

\begin{figure}
  \centering
  \includegraphics[width=0.8\columnwidth]{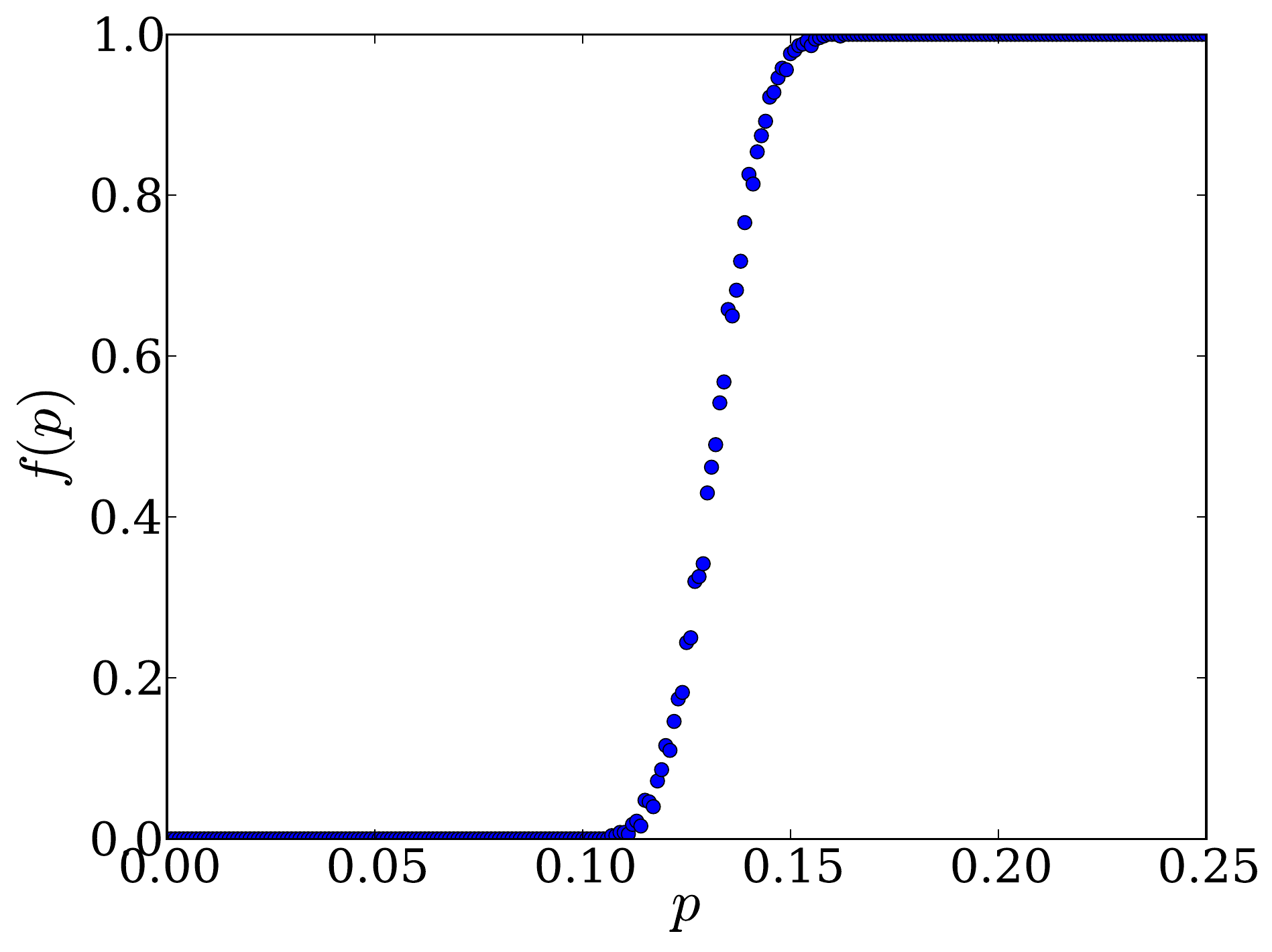}
  \caption{(Colour online) Fraction of desynchronised networks $f(p)$ vs. probability
    of inhibitory links $p$ for 100 realisations of SW networks of $N=100$ and $k=24$ for
    $\sigma=0.3$, $\tau=10$, $b=0.95$.}
  \label{fig:trans_N100_k18_small_ZOOM}
\end{figure}

As an example, Fig.~\ref{fig:trans_N100_k18_small_ZOOM} shows the
transition from stable synchronisation, where the fraction $f$ of
network realisations with unstable synchronisation is zero, to unstable
synchronisation, where $f=1$, as the probability of inhibition $p$ is
increased. This example is for the coupling parameters $\sigma=0.3$
and $\tau=10$, but this transition to inhibition-induced desynchronization
will look the same for all sufficiently large values of $\tau$. A thorough 
investigation of the effects of additional inhibitory links when $\tau$ is 
small will be presented in Sec.~\ref{section:implications}.

\section{Small delay times}
\label{section:small_tau}

If the delay time is not large enough, the rotational symmetry of the MSF no
longer holds. In fact, while reducing $\tau$ one can witness how the
rotational symmetry begins to break down. This is depicted
in Fig.~\ref{fig:msf_symmetry_breakdown}. For a constant coupling
strength of $\sigma=0.3$, the MSFs are numerically calculated for
decreasing delay times. While at $\tau=10$ the MSF still has its
circular form (Fig.~\ref{fig:msf_symmetry_breakdown}(a)), when decreasing $\tau$, the
stable (i.e. dark blue/green) region of the MSF begins to show signs of
deformation. By $\tau=7$ (see
Fig.~\ref{fig:msf_symmetry_breakdown}(b)) the stable region is clearly
larger than the unit circle scaled by $\sigma=0.3$ and has definitely
lost its rotational symmetry. By $\tau=6.5$ (see
Fig.~\ref{fig:msf_symmetry_breakdown}(c)) the stable region has
already split into two disconnected regions. Letting $\tau$ decrease
further, the stable regions become increasingly smaller (see
Fig.~\ref{fig:msf_symmetry_breakdown}(d)). Note that the delay-induced limit cycle
coexists alongside the stable fixed point and whether it is reached or not is
therefore dependent on initial conditions. For $\tau$ less than about
4 (not shown here), the coupling is no longer able to induce the
homoclinic bifurcation that creates the limit cycle 
at all (as discussed in Sec.~\ref{section:model}), in other words, the neurons no longer
oscillate. Instead, all solutions approach the stable fixed point.

\begin{figure}
\centering
  \begin{overpic}[width=0.9\columnwidth]{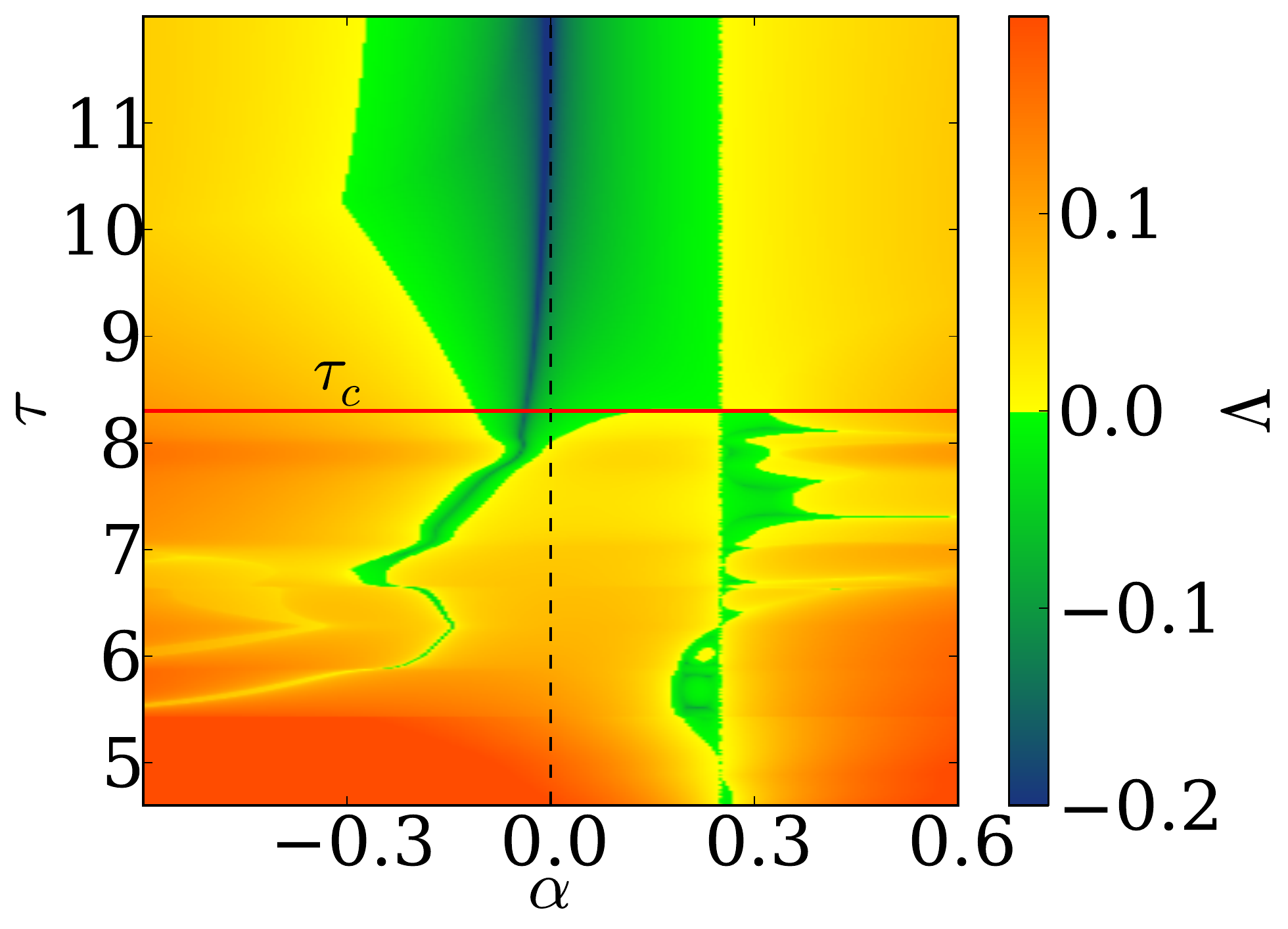}
  \put(14,66){\small(a)}
  \end{overpic}
  \begin{overpic}[width=0.9\columnwidth]{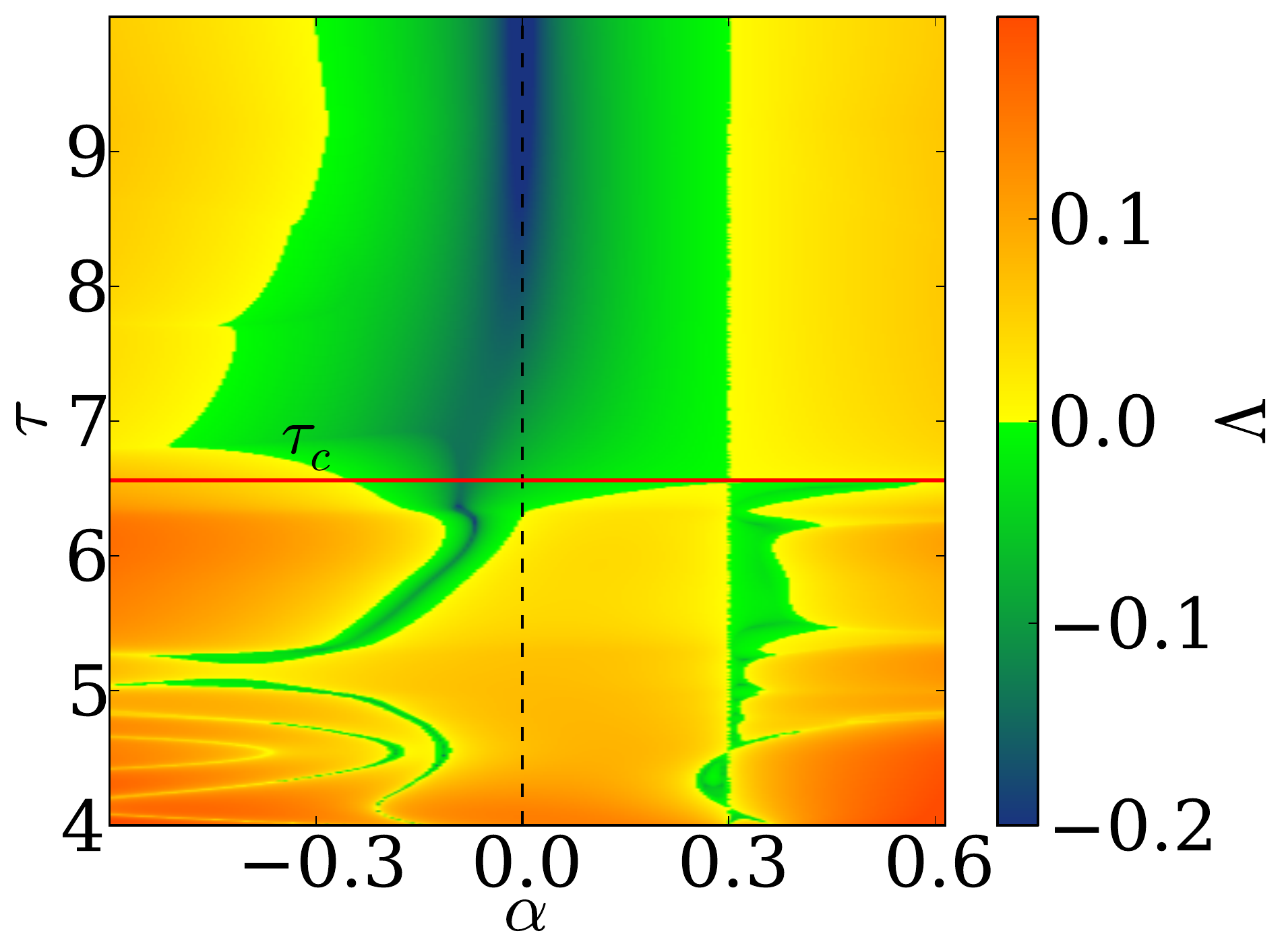}
  \put(12,68){\small(b)}
  \end{overpic}
  \begin{overpic}[width=0.9\columnwidth]{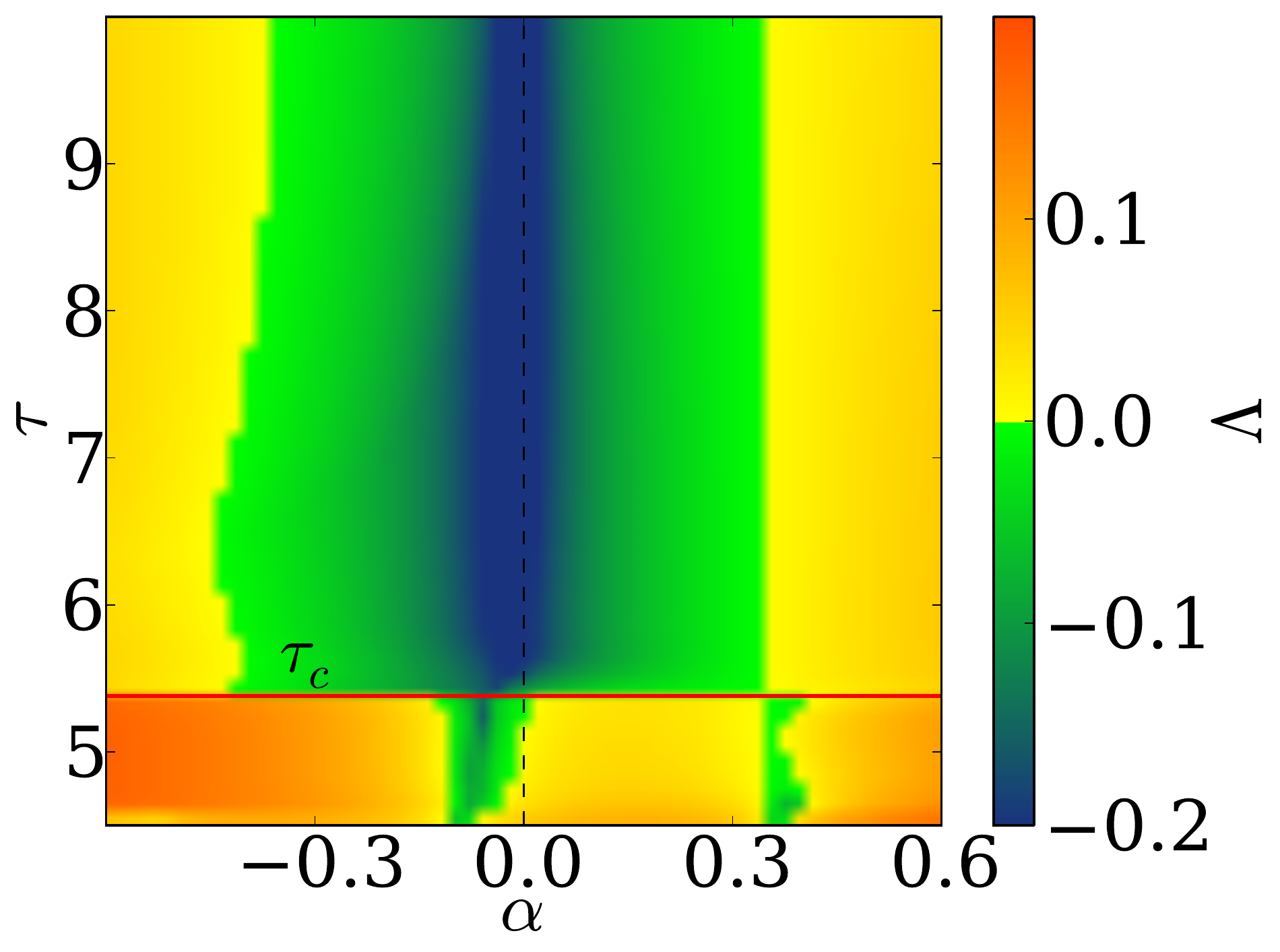}
  \put(12,68){\small(c)}
  \end{overpic}
  \caption{(Colour online) MSF $\Lambda$ for a fixed coupling strength (a)
    $\sigma=0.25$, (b) $\sigma=0.3$ and (c) $\sigma=0.35$ 
    in the plane of the real part $\alpha$ ($\beta=0$) and the delay time $\tau$.   
    The horizontal red lines show the position of
    the critical delay time $\tau_c$. $b=0.95$.}
  \label{fig:real_MSFs}
\end{figure}

It is now obvious that, in this regime of small delay, small changes in $\tau$ can have
a great impact on the stability. Despite the seemingly sudden change
in the MSFs in Fig.~\ref{fig:msf_symmetry_breakdown} between $\tau=7$
and $\tau=6.5$, the evolution of the boundary of stability is in fact
a continuous one (except at a critical value $\tau_c$, which will be
discussed below). This becomes clear after plotting the MSF versus the real part 
$\alpha$ of the eigenvalue (with $\beta=0$) for varying $\tau$. Taking this
one slice of the eigenvalue plane gives a good indication of the
growth and decay of the stable region in the MSF while changing
$\tau$. Figure~\ref{fig:real_MSFs} shows the MSF as a function of the real part 
$\alpha$ ($\beta=0$) and the delay time $\tau$,
with fixed coupling strengths of (a) 0.25, (b) 0.30 and (c) 0.35.
This produces an MSF on an $(\alpha,\tau)$-plane,
which can be used for network topologies with a coupling matrix that
yields real eigenvalues, e.g. symmetric matrices for undirected
networks. In the following we restrict ourselves to undirected networks.

Based on the exemplary coupling strengths in Fig.~\ref{fig:real_MSFs}
it is clear that the stability depends on both $\sigma$ and $\tau$. In
all examples the vertical boundary line at $\alpha=\sigma$,
corresponding to the longitudinal eigenvalue (i.e. $\nu_0=1$, where
$\Lambda=0$), can be easily identified. It separates regimes of stable 
and unstable synchronisation. Another common characteristic
is that the $\tau$-dependent MSFs have a critical delay time $\tau_c$
at which the stable region splits into two separate, disconnected regions. For
values above $\tau_c$ the stable region is found to the left of the
longitudinal eigenvalue; whereas for values below $\tau_c$ there may
be one stable region to the right of the longitudinal eigenvalue
and one to the left. $\tau_c$ seems to be an important value, because
it marks the most significant $\tau$-dependent qualitative change in
the MSF. Ultimately, the MSF can be divided into three regimes of
$\tau$: (i) two or more smaller separate regions of stability exist when
$\tau < \tau_c$; (ii) one large region of stability exists when $\tau >
\tau_c$; (iii) the MSF has one rotationally symmetric region of stability in
the limit of $\tau \rightarrow \infty$ (which holds already in good approximation
if $\tau$ is of the order of the intrinsic oscillation period or larger).

\begin{figure}
\centering
  \begin{overpic}[width=0.526\columnwidth]{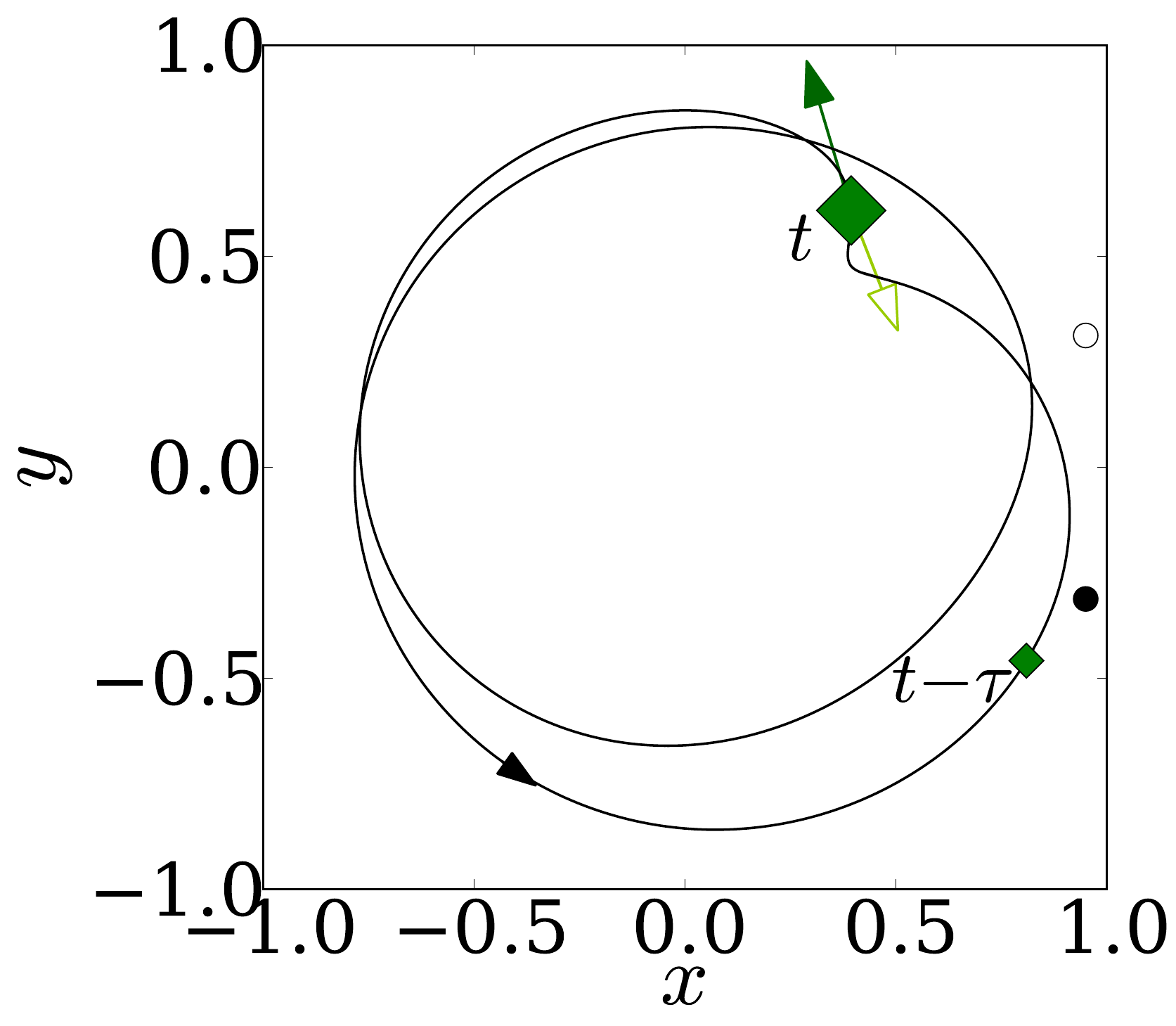}
  \put(24,74){\small(a)}
  \end{overpic}
  \begin{overpic}[width=0.454\columnwidth]{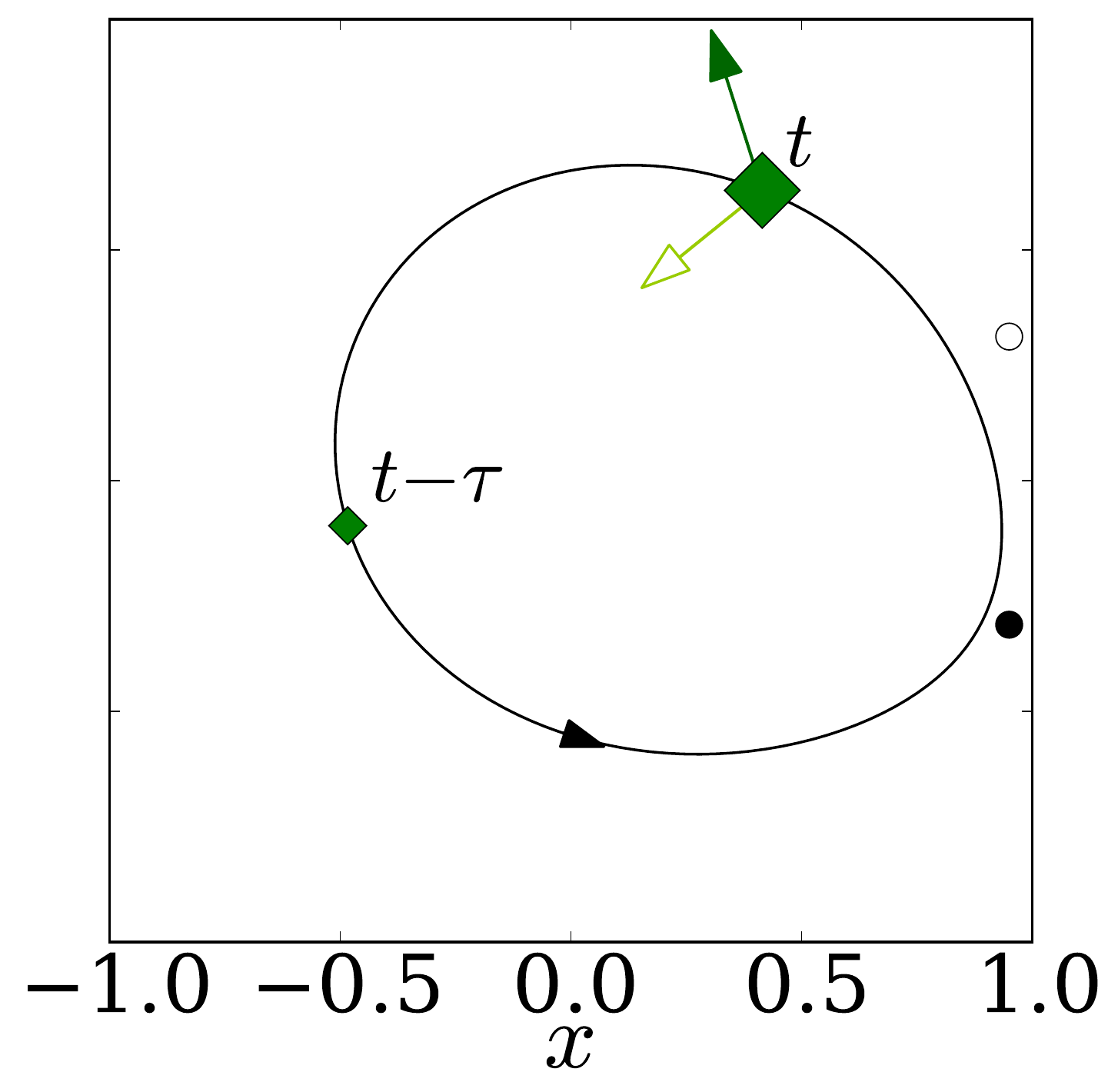}
  \put(12,85){\small(b)}
  \end{overpic}
  \caption{
(Colour online) The position in phase space of one element representing the
    synchronised dynamics of a network (large green diamond) is shown
    relative to its delayed position (small green diamond). The black
    circle and the white circle represent the stable and saddle point,
    respectively. For small delay times such as (a) $\tau=5.35$ a kink
    may be the result of the relative positioning. For larger delay
    times such as (b) $\tau=6$ this does not occur. The movement of
    the phase point has a local component (dark green arrow) and a coupling
    component (light green arrow). The coupling component is directed
    towards the phase point at time $t-\tau$ (small green circle).
    $\sigma=0.35$, $b=0.95$.}
  \label{fig:tau10_535e-2_sigma35e-2_verlauf}
\end{figure}

Clearly, the topological criteria for stable synchronisation will be
very different depending on whether the network is coupled with $\tau
< \tau_c$ or $\tau > \tau_c$. Observing the change in the limit cycles
while varying $\tau$ offers the explanation for the qualitative change
either side of $\tau_c$ in the MSF. For smaller $\tau$ values, the
trajectory of the limit cycle is sensitive to changes in the coupling
parameters, whereas, for larger $\tau$ values, the trajectory changes
very gradually and approaches an elliptical form. The most predominant
feature of the changing limit cycle is a large \textquotedblleft
kink\textquotedblright~for values below $\tau_c$.
Figure~\ref{fig:tau10_535e-2_sigma35e-2_verlauf}(a) shows such a kink in
the limit cycle induced by the coupling parameters $\sigma=0.35$ and
$\tau=5.35$: upon bypassing the stable node the limit cycle deviates a
little towards the unstable focus, located at the origin, before
suddenly changing its trajectory back towards the saddle point. The
kink has a large effect on the dynamics of synchronised nodes and is
an important source of instability. Simulations have shown that it is
at the kink that synchronised nodes begin to diverge from one another
following any small perturbations. For larger $\tau$ the kink
diminishes and disappears (see
Fig.~\ref{fig:tau10_535e-2_sigma35e-2_verlauf}(b)).

Let us explain the cause for the kink in greater detail. The large green 
diamond in Fig.~\ref{fig:tau10_535e-2_sigma35e-2_verlauf}(a) is the 
position of a node that represents the dynamics of a
synchronised network in the phase space. The small green diamond is its
delayed phase point, to which it is coupled, and represents the delayed
dynamics of the same synchronised network (i.e. the small green diamond
at time $t$ is the large green diamond at time $t-\tau$). A node coupled
with a delayed node is attracted towards the direction of this delayed
node (cf. Fig.~\ref{fig:tau10_535e-2_sigma35e-2_verlauf}(b)). As the node passes between
the stable fixed point and the saddle point, it is slowed down, allowing its delayed
node to \textquotedblleft catch up\textquotedblright~to it. For
smaller $\tau$ this has the effect that the node is attracted in a
clockwise direction towards its delayed node and gets slowed down (cf.
Fig.~\ref{fig:tau10_535e-2_sigma35e-2_verlauf}(a)). This is the cause
for the kink. Whereas for larger $\tau$, the (small green) delayed
node is always more than half a rotation behind, meaning that the
(large green) node is attracted towards it in an anti-clockwise
direction. This gives the limit cycle a form that is roughly circular
as seen in Fig.~\ref{fig:tau10_535e-2_sigma35e-2_verlauf}(b).

As long as the kink is present in the limit cycle, a synchronised
network will generally have very different topological conditions in
order to be stable. 
Even when the limit cycle has a very
predominant (sharp) kink, it will often be possible to find a class of
networks that will still show stable synchronisation.

Note that in Fig.~\ref{fig:real_MSFs} one
can see how the most stable part of the stable region (i.e. where
$\Lambda$ is at its smallest) shifts away from $\alpha=0$ as $\tau$
becomes smaller. Simulations have shown that this happens while the angle
 of the kink increases (making the kink ``sharper'').
A possible interpretation could be as follows. 
Each potential eigenvalue of the coupling matrix in a MSF is associated with one or more
eigenvectors in the $N$-dimensional space.
These eigenvectors are all orthogonal to the vector $(1, 1, 1, ..., 1)$ that represents 
motion in the SM.
This would suggest that the increasing angle of the kink corresponds
to a shifting of the most stable direction of perturbation
transverse to the SM.

\section{Implications for complex networks with a small delay coupling
  time}
\label{section:implications}

An obvious observation is that networks with purely excitatory coupling, which are
always stable for large delay times as mentioned at the end of
Sec.~\ref{section:MSF}, may not show stable synchronisation for $\tau < \tau_c$. It was
explained in Sec.~\ref{section:MSF} that increasing the probability $p$ of
inhibitory links in the network can be a factor leading to
unstable synchronisation. This occurs because a larger probability of
inhibition can push a part of the eigenvalue spectrum of any network
beyond the longitudinal eigenvalue at $\alpha=\sigma$ (because Gershgorin's circle
theorem no longer holds) and into the
unstable region of the MSF. Now, for smaller $\tau$, there may be a
pocket of stability to the right of the longitudinal eigenvalue, so
that increasing inhibition can make the otherwise unstable
synchronisation of a network stable.


This means that the transitions between stable and unstable synchronisation as a function of
the probability of inhibition $p$ discussed in Ref.~\cite{LEH11} are no longer
valid when $\tau$ is small. The transitions are now
sensitive to the coupling parameters, not just the network parameters
$N$ and $k$. Due to the multiple regions of stability, eigenvalues may
wander in and out of stable regions, while increasing the probability
of inhibition $p$. For large $\tau$, increasing $p$ in the SW network
model only results in one transition where the fraction of
desynchronising networks (i.e. networks with unstable synchronisation)
$f(p)$ switches from 0 to 1 (cf.~Fig.~\ref{fig:trans_N100_k18_small_ZOOM}). 
For small $\tau$, it is possible that $f(p)$
jumps back to 0, before increasing again to 1. This will occur if
there is a separate region of stability to the right side in the MSF that is
large enough that all the eigenvalues lying over there can fit inside.


The observation of multiple transitions between stable and unstable
synchronisation can be explained by looking at the eigenvalue spectra
for SW networks for various $p$ values. As discussed above in relation
to the MSF method, each network topology has a coupling matrix
$\mathbf{G}$, the eigenvalues from which can be used to determine the
stability of the network's synchronised state. Because the
\textquotedblleft short-cuts\textquotedblright~in the SW network are
randomly introduced, each network with specific $N$, $k$ and $p$ values
may have many realisations. By calculating the eigenvalues for a large
number of realisations of the SW network with certain parameters (i.e.
given $N$, $k$ and $p$), one can find the bounds for the eigenvalue
spectra.

\begin{figure}
  \centering
  \includegraphics[width=0.8\columnwidth]{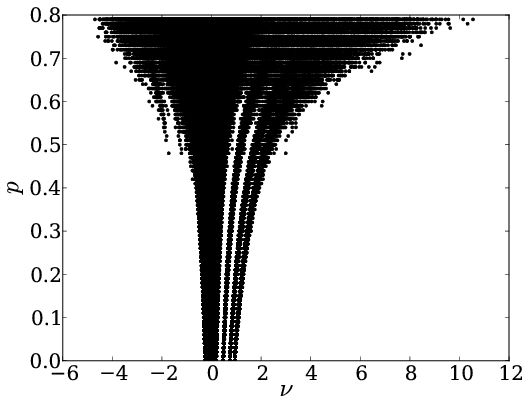}
  \caption{Spectra of eigenvalues $\nu$ of undirected small world
    networks with $N=200$ and $k=20$ and varying probability $p$ of inhibitory links.
500 realisations of the eigenvalue spectrum are plotted for each value of $p$.}
  \label{fig:eig_spectrum_small-world_N200k20_wander}
\end{figure}

Figure~\ref{fig:eig_spectrum_small-world_N200k20_wander} displays the
superimposed eigenvalue spectra of 500 realisations for SW networks with
parameters $N=200$ and $k=20$ (regular ring with excitatory coupling of 
$k$ nearest neighbours on either side) in dependence on the probability 
of inhibition $p$. The longitudinal eigenvalues located at $\nu_1=1$ have 
been removed. One can see that the bounds for possible
eigenvalues shift depending on the value of $p$. One can also see
how the spectra begin to increasingly resemble the semicircular
distribution~\cite{ALB02a} of a random network for larger $p$ values,
where the networks have lost their SW properties.

\begin{figure*}
  \begin{overpic}[width=0.33\textwidth]{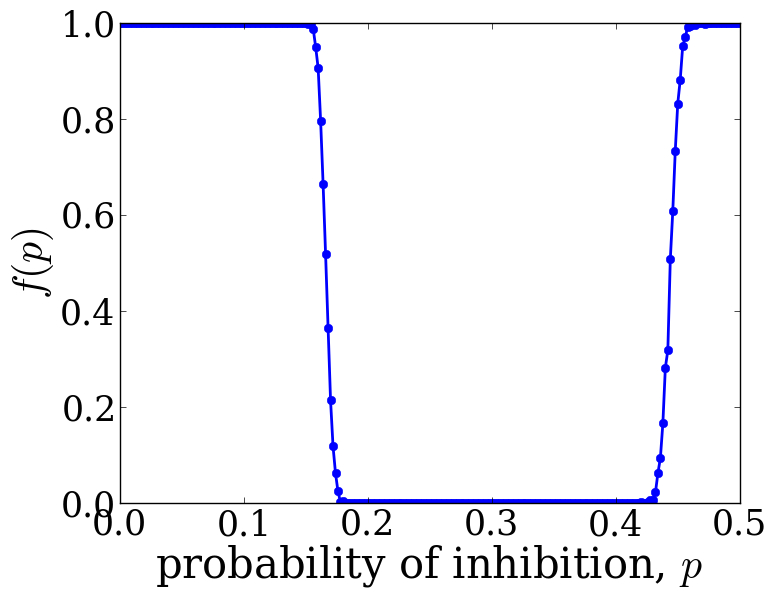}
  \put(18,65){\tiny(a)}
  \end{overpic}
  \begin{overpic}[width=0.33\textwidth]{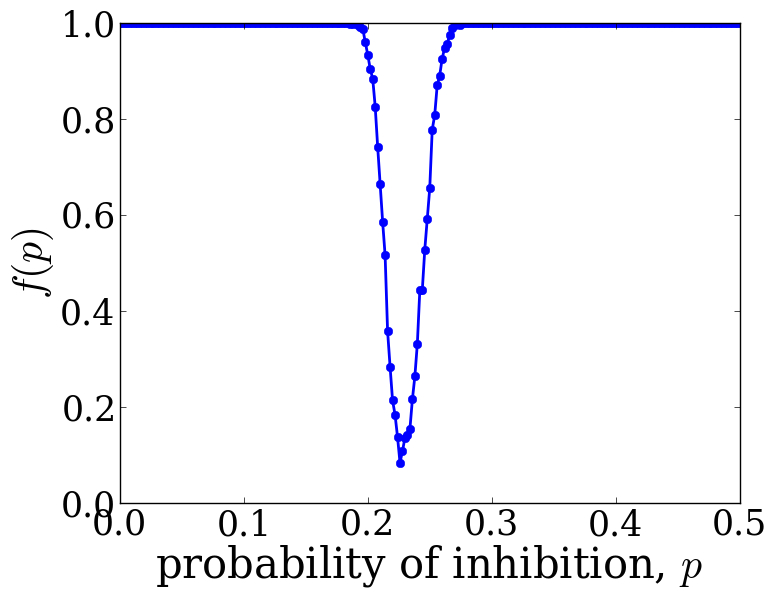}
  \put(18,65){\tiny(c)}
  \end{overpic}
  \begin{overpic}[width=0.33\textwidth]{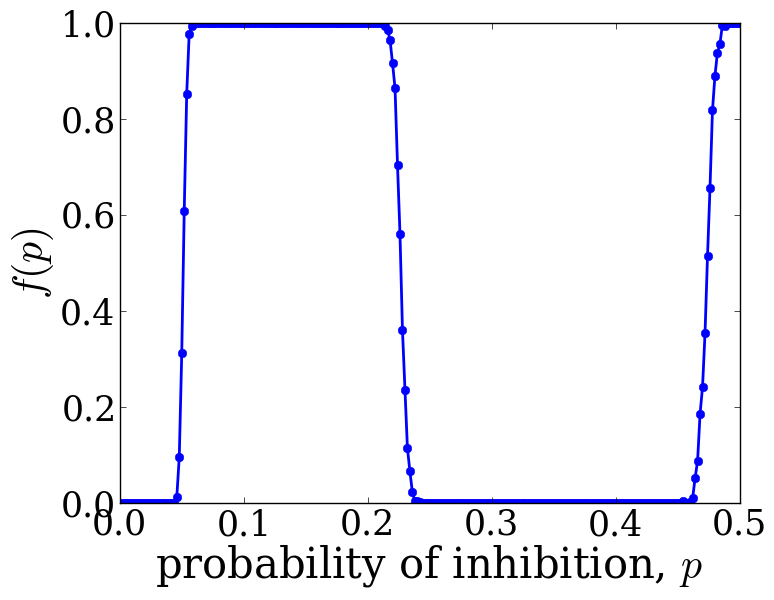}
  \put(18,65){\tiny(e)}
  \end{overpic}
  \begin{overpic}[width=0.33\textwidth]{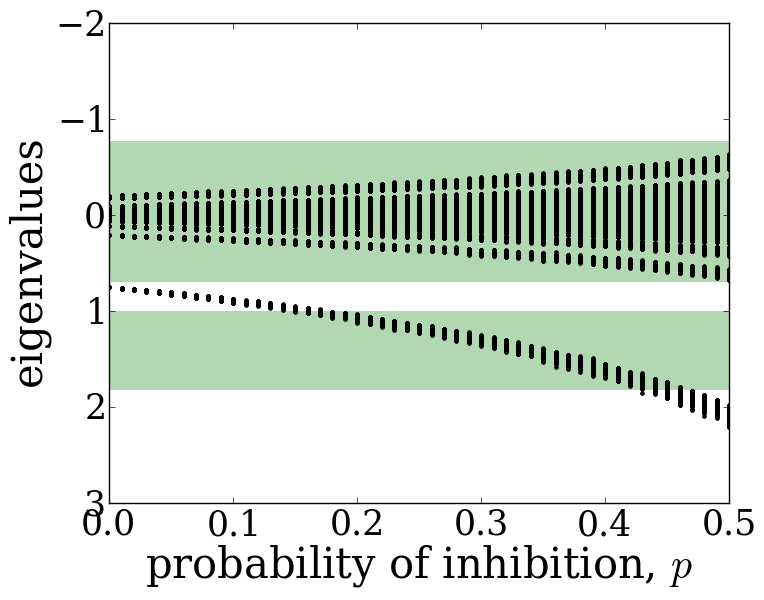}
  \put(18,65){\tiny(b)}
  \end{overpic}
  \begin{overpic}[width=0.33\textwidth]{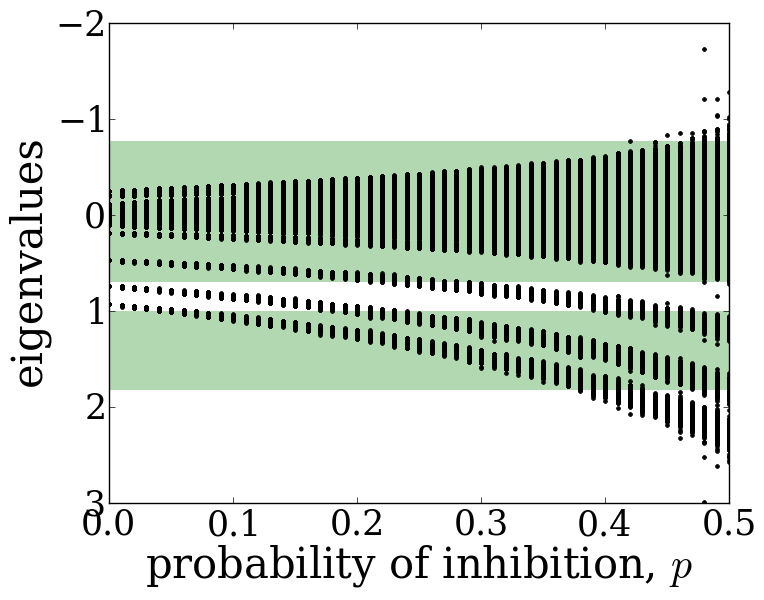}
  \put(18,65){\tiny(d)}
  \end{overpic}
  \begin{overpic}[width=0.33\textwidth]{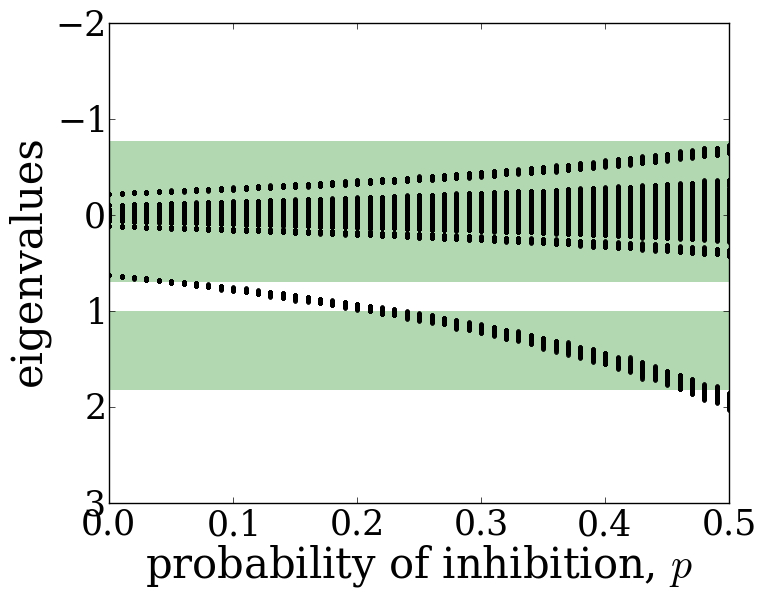}
  \put(18,65){\tiny(f)}
  \end{overpic}
  \caption{(Colour online) Fraction of desynchronized networks $f$ in dependence of
    the probability of additional inhibitory links $p$ for 500 realisations of networks of
    $N=200$ with (a) $k=40$, (c) $k=20$, and (e) $k=50$ with
    $\sigma=0.3$ and $\tau=6.5$. Corresponding eigenvalue spectra for
    (b) $k=40$, (d) $k=20$, and (f) $k=50$ with $100$ realisations for
    each $p$ value. Here the green shaded regions represent the stable
    regions of the real part of the MSF. $b=0.95$.}
  \label{fig:trans_N200_k20_k40_k50_small}
\end{figure*}

The eigenvalue spectra bridge the gap between observations of the MSF
(i.e. the dependence of the stability of synchronisation on the
eigenvalues) and what is seen in these transitions (i.e. the
dependence of stability on the network topology). Increasing $p$
allows isolated eigenvalues to increase in value and, in terms of the
MSF, shift their locations further to the right in the $(\alpha,
\beta)$-plane. This is visualised in Fig.~\ref{fig:trans_N200_k20_k40_k50_small}.
Figures~\ref{fig:trans_N200_k20_k40_k50_small}(b), (d) and (f)
show the eigenvalue spectra for SW networks with $N=200$ elements and
$k=40$, 20, and 50, respectively.

Figures~\ref{fig:trans_N200_k20_k40_k50_small}(a), (c) and (e) show the
corresponding fraction of desynchronised networks $f$ in dependence on the
probability of additional inhibitory links $p$. Consider, for instance, SW networks
with parameters $N=200$ and $k=40$. In
Fig.~\ref{fig:trans_N200_k20_k40_k50_small}(a) $f(p)$ is shown for the
exemplary coupling parameters $\sigma=0.3$ and $\tau=6.5$ with a
corresponding plot of eigenvalues in
Fig.~\ref{fig:trans_N200_k20_k40_k50_small}(b). Note that the grey
(green) shaded regions represent the stable regions of the line
$\beta=0$ on the $(\alpha,\beta)$-plane of the MSF for $\sigma=0.3$
and $\tau=6.5$ (cf. Fig.~\ref{fig:msf_symmetry_breakdown}(c)). For
coupling parameters where this region of stability is not large
enough, $f$ may briefly dip below $1$ without decreasing to $0$,
because only some realisations may have eigenvalues that fit inside
the stable region. In Fig.~\ref{fig:trans_N200_k20_k40_k50_small}(c)
where $k=20$, $f$ dips down to $0.14$, because at most $14\%$ of the
realisations show stable synchronisation (i.e. all the eigenvalues are
inside the stable region). If the distance between the larger isolated
eigenvalues matches the distance between stable regions (which is
almost the case in Fig.~\ref{fig:trans_N200_k20_k40_k50_small}(d)),
then the transition curve only just touches the $f=0$ axis at some value of $p$ before
increasing back to $f=1$.
Figure~\ref{fig:trans_N200_k20_k40_k50_small}(e) is an example where a
further transition is possible, because not only do all eigenvalues
begin in stable regions at $p=0$, but there is another regime of $p$
where all eigenvalues fit inside stable regions. Note that the length
of the transition from $f=0$ to $f=1$ is actually a measure of the
variance of the isolated eigenvalues for an ensemble of realisations.
For example, the variance of the isolated eigenvalues decreases as $N$
is increased, so that the length of transition will be shorter in larger networks,
and the transition becomes sharper.

\begin{figure}
\centering
  \begin{overpic}[width=0.49\columnwidth]{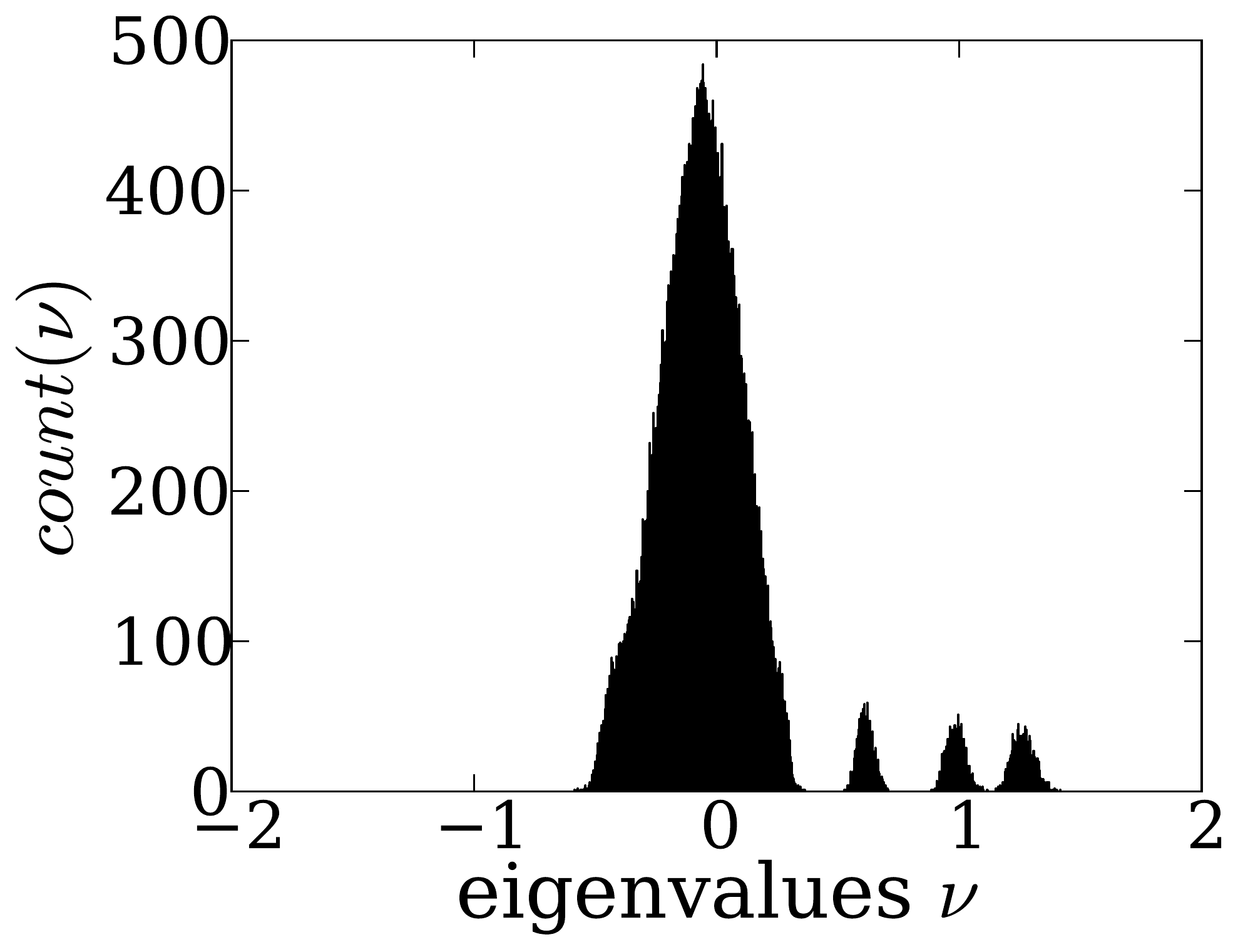}
  \put(0,68){\tiny(a)}
  \end{overpic}
  \begin{overpic}[width=0.49\columnwidth]{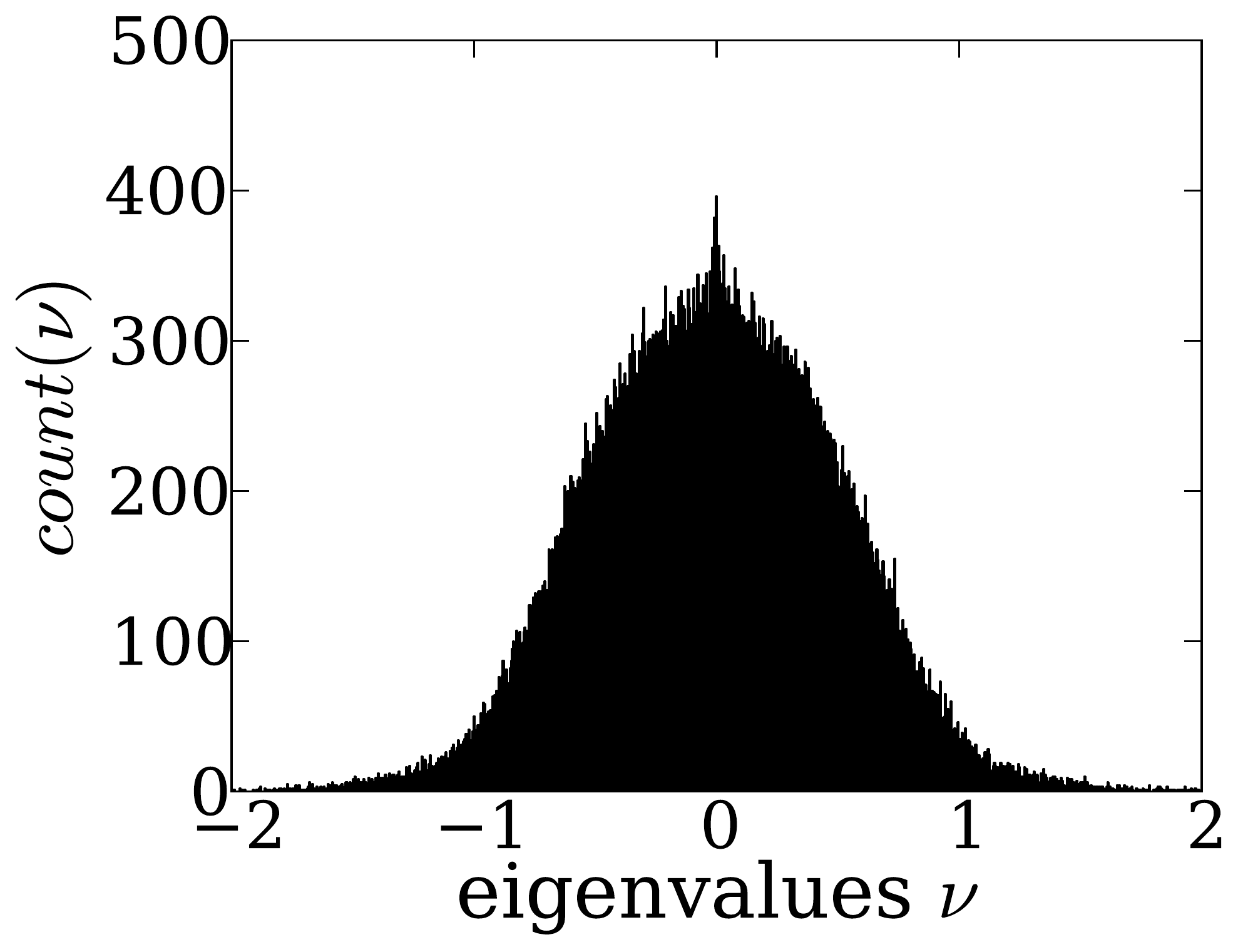}
  \put(0,68){\tiny(b)}
  \end{overpic}
  \caption{
Histograms of eigenvalue spectra for $1000$ realisations of
    (a) SW networks and (b) random networks with the same number of
    nodes, the same number of excitatory links and the same
    probability of inhibitory links. $N=100$, $k=10$, $p=0.2$ and number of bins is $1000$.}
  \label{fig:eig_spectrum_small-world_random_N100k10p0.2}
\end{figure}

\begin{figure}
  \centering
  \includegraphics[width=0.8\columnwidth]{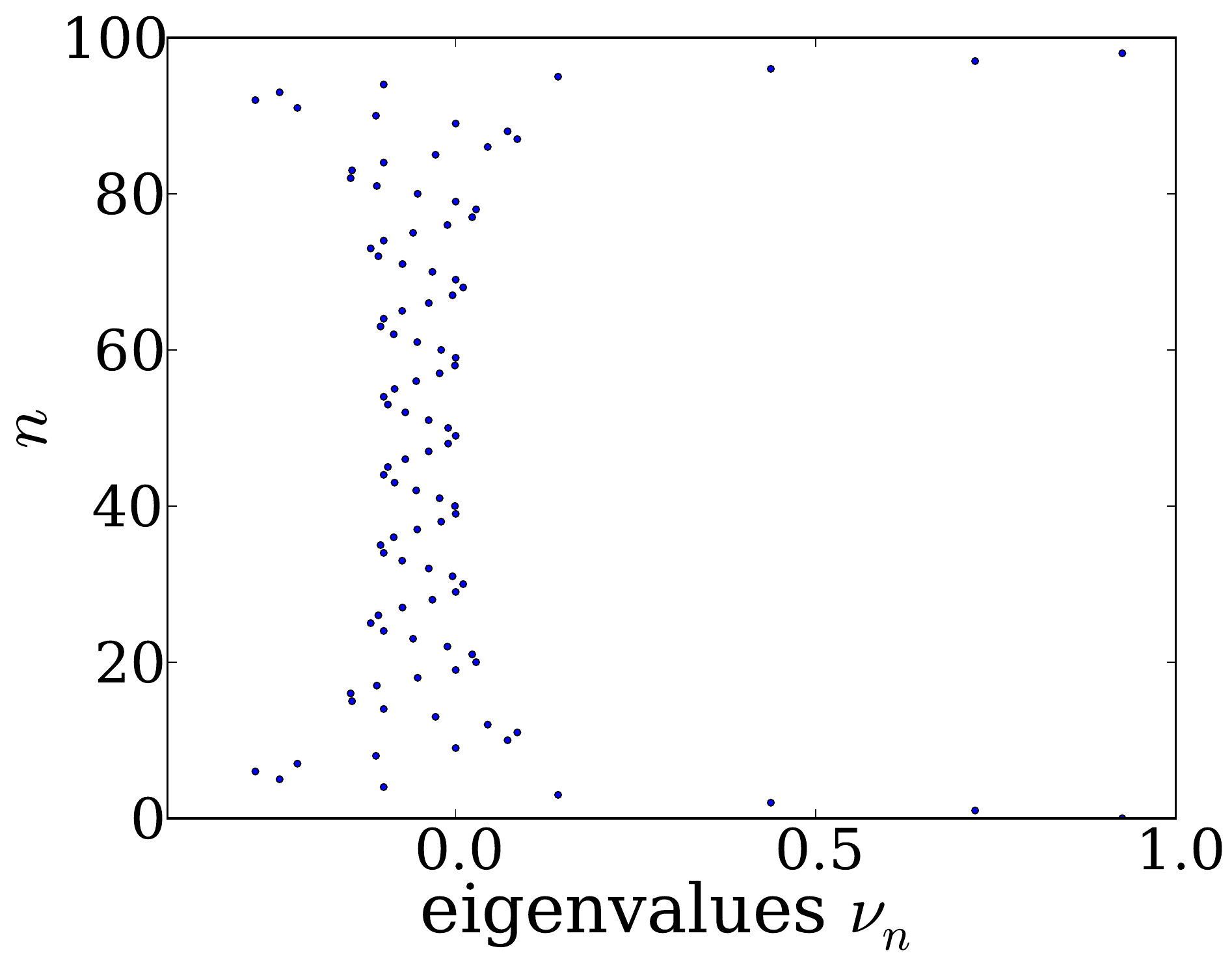}
  \caption{
The eigenvalue spectrum for a regular network with $N=100$
    and $k=10$. $\nu_n$ is the value of the $n^{th}$ eigenvalue.}
  \label{fig:eig_spectrum_regular_N100k10}
\end{figure}

When constructing a histogram of the eigenspectra for a particular
value of $p$ -- see
Fig~\ref{fig:eig_spectrum_small-world_random_N100k10p0.2}(a) -- the
larger isolated eigenvalues seen to the right, e.g. in
Fig.~\ref{fig:trans_N200_k20_k40_k50_small}(d), result in small
peaks of eigenvalues.
For each realisation of many SW networks there are isolated
eigenvalues that are larger than most other eigenvalues in the
spectrum. The small peaks of eigenvalues result from perturbations in
isolated eigenvalues. The spectrum of a regular ring network (i.e. a
SW network with $p=0$) can be found analytically using the graph's
symmetry operations \cite{FAR01} and is given by
\begin{eqnarray}
  \label{eq:eig_regular}
  \nu_l &=& \frac{1}{k} \sum\limits_{j=1}^k \cos\left(2\pi j \frac{l}{N}\right) \nonumber \\
  &=& \frac{1}{k} \left(\frac{\cos\left(k\pi\frac{l}{N}\right) \sin\left((k+1) \pi \frac{l}{N}\right)}{\sin\left(\pi \frac{l}{N}\right)} -1 \right),
\end{eqnarray}
where $l=1,...,N-1$. Figure~\ref{fig:eig_spectrum_regular_N100k10}
shows this eigenspectrum for the $p=0$ case.
As $p$ is increased these eigenvalues will be slightly perturbed (in a
random manner) by the changing network structure, of which many
realisations are possible. In
Fig.~\ref{fig:eig_spectrum_regular_N100k10} one can see the isolated
eigenvalues to the right-hand side of the plot that eventually evolve into the
smaller side peaks of eigenvalues in histograms for larger $p$ (see
Fig~\ref{fig:eig_spectrum_small-world_random_N100k10p0.2}(a)). These
peaks show the distribution of the eigenvalue under the influence of
the random nature of the SW ``short-cuts'' creation process -- which is
why the smaller eigenvalue peaks can be approximated by a normal
distribution~\footnote{Actually, because the multiplicity of the above
mentioned isolated eigenvalues at $p=0$ is $2$, the smaller eigenvalue
peaks are two normal distributions that, at least for small $p$ values,
overlap each other to a large extent}. 
This is furthermore the reason why the desynchronisation transitions observed in
Ref.~\cite{LEH11} and Figs.~\ref{fig:trans_N200_k20_k40_k50_small}(a),
(c) and (e) look like the cumulative (integrated) distribution function. They
reflect how increasing $p$ brings the area of the small eigenvalue
peak accumulatively into the unstable region of the MSF.

To explain why the peaks wander with increasing $p$, consider not normalising the rows
of matrix $\mathbf{G}$, so that the row sums are not necessarily equal
to $1$. Then the location of the longitudinal eigenvalue decreases
with $p$, because it is equal to the average row sum of $\mathbf{G}$
which is equal to $2k(1-p)$. The locations of the other peaks
increase slightly to maintain the eigenvalue sum of zero (a result
of the trace of $\mathbf{G}$ being zero, since there is no
self-feedback coupling), although for large $N$ this has only a small
effect. Thus, scaling the eigenvalues so that the longitudinal
eigenvalue is always at $1$ means that the transversal eigenvalues are
multiplied by $\frac{1}{2k(1-p)}$, so that they appear to increase
with $p$, as seen in Figs.~\ref{fig:trans_N200_k20_k40_k50_small}(b), (d) and (f).

The case of random Erd\H{o}s-R\'{e}nyi networks 
\cite{ERD59} is different. At $p=0$ there are no larger gaps between
the eigenvalues, such as those for SW networks at $p=0$. If a random
network is fully connected (i.e. is one single component) then the
distribution of its eigenvalues at $p=0$ are confined to a
semi-circle, given by the semicircular distribution
\begin{equation}
  \label{eq:eig_random_p0}
  \rho(\nu) = 
  \begin{cases}
    \frac{\sqrt{4Nq(1-q)-[\frac{\nu}{q(N-1)}]^2}}{2\pi Nq(1-q)} & \textnormal{if } \| \nu \| < 2 \frac{\sqrt{Nq(1-q)}}{q(N-1)} \\
    0 & \textnormal{otherwise} , 
  \end{cases}
\end{equation}
where $q$ is the probability of
excitatory links~\cite{ALB02a}. While there are initially no gaps in
the eigenvalue distribution at $p=0$, adding inhibitory links to the
network (i.e. increasing $p$) does not lead to the creation of any
gaps like those seen for SW networks. To illustrate this point,
Fig.~\ref{fig:eig_spectrum_small-world_random_N100k10p0.2}(b) shows
the eigenvalue distribution for many realisations of
Erd\H{o}s-R\'{e}nyi random networks with the same number of nodes, as
well as the same number of excitatory links and the same probability
of inhibitory links. As a result, multiple transitions between stable
and unstable synchronisation will not occur.

One can, however, use Eq.~(\ref{eq:eig_random_p0}) to find parameters
$N$ and $q$ such that the eigenvalue distribution is confined to a
stable region of the MSF at $p=0$, as long as there is a stable region
centred around $\nu=0$ (for example, see
Fig.~\ref{fig:msf_symmetry_breakdown}(c)). This allows for a
desynchronisation transition with increasing $p$, such as 
in Fig.~\ref{fig:trans_N100_k18_small_ZOOM}.

\section{Small-world networks of Stuart-Landau oscillators}
\label{sec:stuart-landau}

The appearance of multiple transitions between synchronisation
and desynchronisation is based on the interplay of the peculiar spectral
properties of the small-world topology and the unconnected stable
regions that show up in the MSF for the excitable SNIPER model for
small delay times.

In this Section we show that such unconnected stable regions may also
appear in the MSF of a purely oscillatory system. We consider the
Stuart-Landau oscillator, which is the normal form of any oscillatory
system near a supercritical Hopf bifurcation. The dynamics of the
Stuart-Landau oscillator is governed by a complex variable $z \in
\mathbb{C}$, such that we have $\mathbf{x}_i=z_i$ in
Eq.~\eqref{eq:neteqns}. The local dynamics of the Stuart-Landau
oscillator is given by
\begin{equation}
  \label{eq:1}
  f(z) = (\lambda+i\omega - \left|z\right|^2)z.
\end{equation}
For $\lambda>0$, the uncoupled system exhibits self-sustained limit cycle
oscillations with radius $r_0=\sqrt{\lambda}$ and frequency $\omega$.
The coupled system shows -- depending on coupling parameters and
topology -- a variety of synchronised and desynchronised solutions,
including amplitude death and cluster synchronisation \cite{CHO09,CHO11}.

Here we focus on complete (zero-lag) synchronization.
\begin{figure}
  \begin{overpic}[width=0.46\columnwidth]{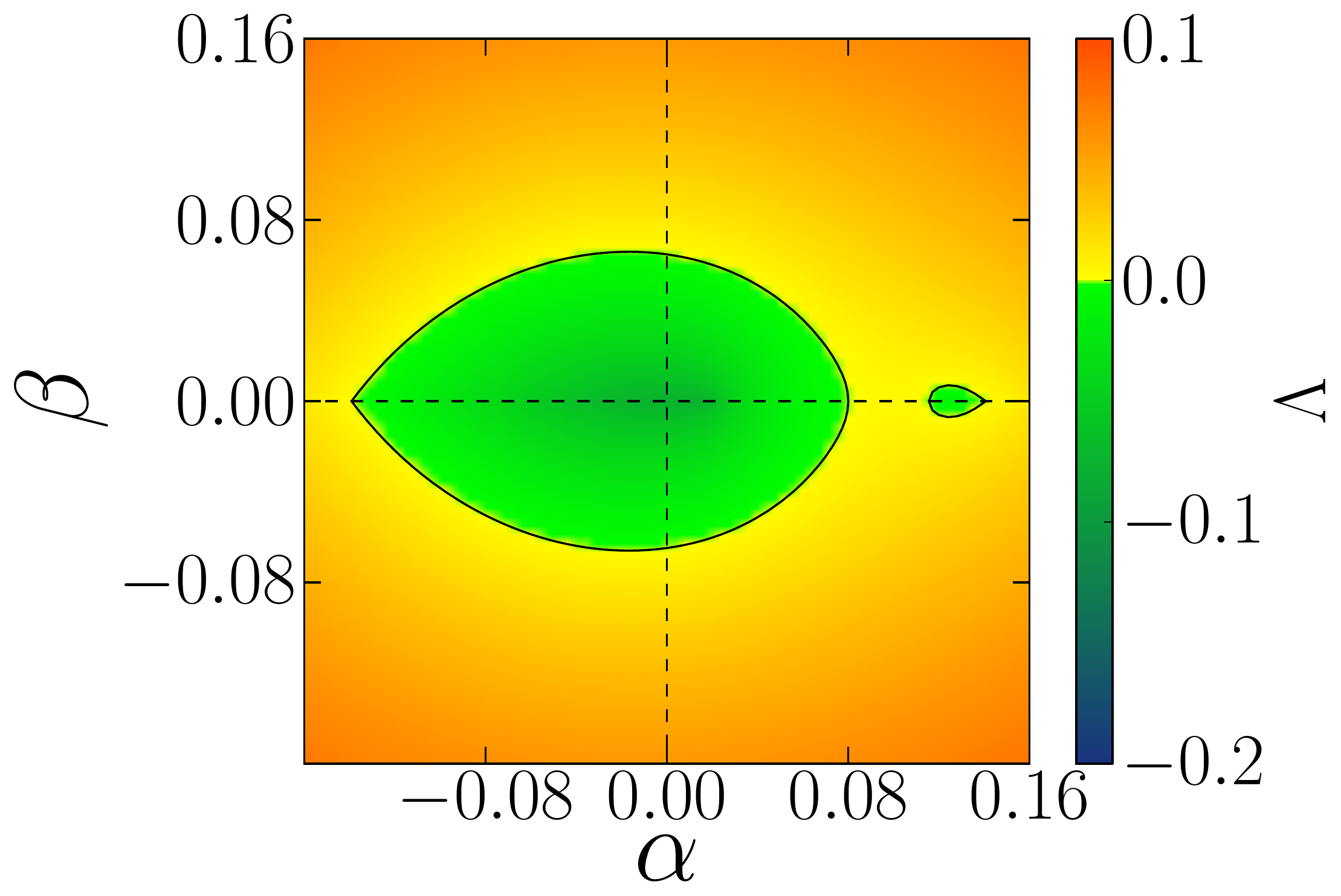}
  \put(25,58){\tiny(a)}
  \end{overpic}
  \hspace*{0.05\columnwidth}
  \begin{overpic}[width=0.46\columnwidth]{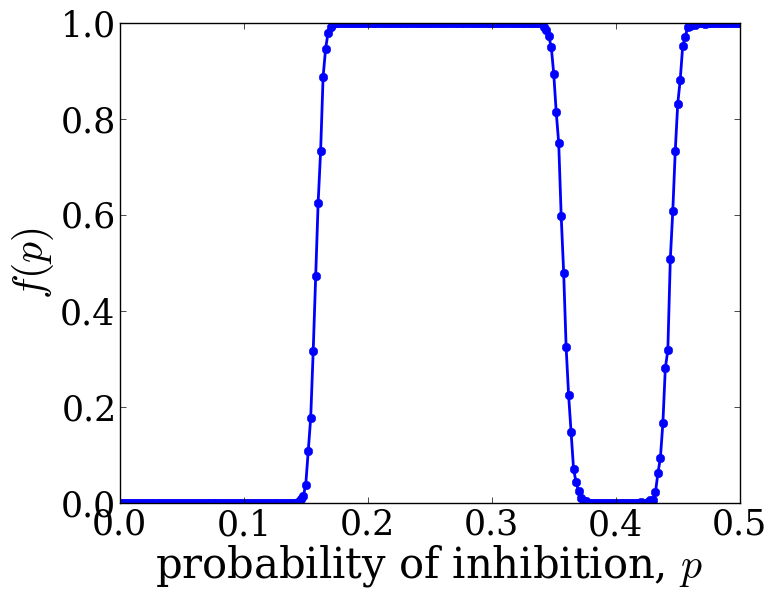}
    \put(18,65){\tiny(c)}
  \end{overpic}
  \begin{overpic}[width=0.46\columnwidth]{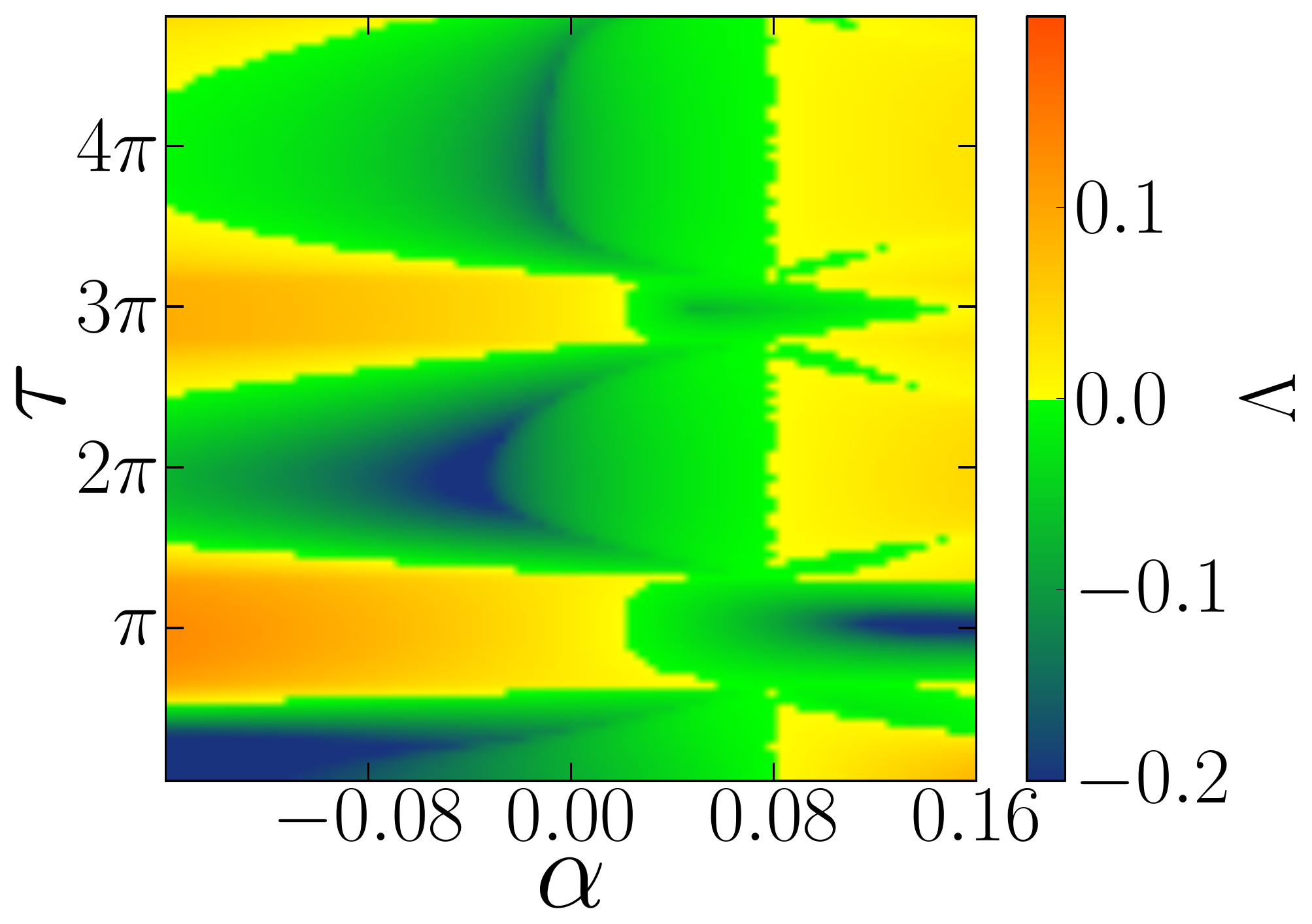}
  \put(15,63){\tiny(b)}
  \end{overpic}
  \hspace*{0.05\columnwidth}
  \begin{overpic}[width=0.46\columnwidth]{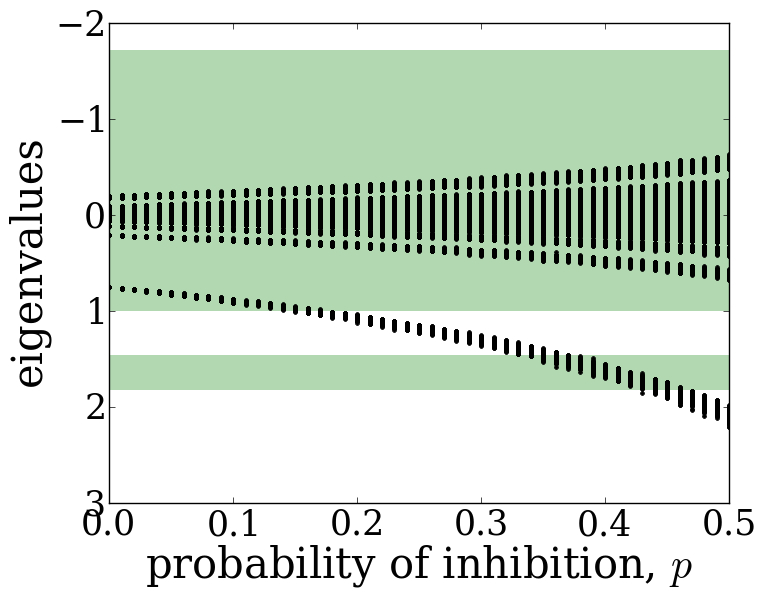}
    \put(18,65){\tiny(d)}
  \end{overpic}
  \caption{
(Colour online) Network of Stuart-Landau oscillators according to
    Eqs.~\eqref{eq:neteqns} and~\eqref{eq:1}: (a) MSF in the
    $(\alpha,\beta)$ plane for $\tau=3\pi/2$. (b)
    MSF in the $(\alpha,\tau)$ plane for
    $\beta=0$. (c) Fraction of desynchronized networks $f$ in
    dependence of the probability of additional inhibitory links $p$
    for 500 realisations of networks of $N=200$ and $k=40$, $\tau=3\pi/2$. 
    (d) Corresponding eigenvalue spectra, where the green shaded areas represent stable
    regions. Other parameters: $\lambda=0.1$, $\omega=1$,
    $\sigma=0.08$.}
  \label{fig:sl_msf}
\end{figure}
Fig.~\ref{fig:sl_msf}(a) shows the MSF for sets of parameters that allow
for unconnected stable regions in the complex plane of the eigenvalues 
$(\alpha,\beta)$. As
shown in Fig.~\ref{fig:sl_msf}(b) a connected stability region occurs for integer
multiples of $\tau=2\pi/\omega=2\pi$, while unconnected stability regions -- which
can lead to multiple phase transitions in SW networks -- occur in
between.

For the parameters used in Fig.~\ref{fig:sl_msf}(a),
Fig.~\ref{fig:sl_msf}(c) shows the fraction of desynchronised networks
for the modified small-world model as in
Sec.~\ref{section:implications}. Figure~\ref{fig:sl_msf}(d) shows the
eigenspectra for networks of $N=200$ elements, starting from a one-dimensional 
ring with excitatory coupling to $k=40$ neighbours to either side. 
Increasing the probability of additional inhibitory
links $p$ broadens the spectra and leads to a second regime of stable
synchronisation matching the transitions in Fig.~\ref{fig:sl_msf}(c).

\section{Conclusion}
\label{conclusion}

We have investigated transitions between synchronisation and desynchronisation in complex
networks of delay-coupled excitable elements of type I, induced by varying the balance between excitatory and inhibitory couplings in a small-world topology. 
In our analysis we have used the master stability function approach.
For large delay times it seems that both type-I neurons,
as modelled here, and type-II neurons, as modelled in
Ref.~\cite{LEH11}, must fulfill similar topological conditions in the network 
to allow for
a stable synchronised state. This is different when considering small
delay times. For a range of small coupling strengths and small delay
times we have found novel multiple transitions between synchronisation
and desynchronisation, when the fraction of inhibitory links is increased.
Unlike the Erd\H{o}s-R\'{e}nyi random network, a small world model for complex networks
with regular excitatory couplings and random inhibitory shortcuts has eigenvalue spectra with gaps between the
larger eigenvalues, so that histograms of many realisations reveal
isolated peaks of possible eigenvalues. It was shown that, because of
this, synchronised small world networks can go through multiple transitions of
stability in dependence on the probability of inhibitory short-cuts.
Our results are valid beyond the special model of type-I excitability, as we have demonstrated using the Stuart-Landau oscillator.

This work was supported by the DFG in the framework of the SFB 910. PH
acknowledges support by the BMBF (grant no.~01GQ1001B). 


\begin{thebibliography}{39}

\bibitem{ALB02a}
R.~Albert, A.L. Barab\'asi, Rev. Mod. Phys. \textbf{74}, 47 (2002)

\bibitem{NEW03}
M.E.J. Newman, SIAM Review \textbf{45}, 167 (2003)

\bibitem{BOC06a}
S.~Boccaletti, V.~Latora, Y.~Moreno, M.~Chavez, D.U. Hwang, Physics Reports
  \textbf{424}, 175 (2006)

\bibitem{TIM02a}
M.~Timme, F.~{Wolf}, T.~Geisel, Phys. Rev. Lett. \textbf{89}, 258701 (2002)

\bibitem{STE06a}
A.J. Steele, M.~Tinsley, K.~Showalter, Chaos \textbf{16}, 015110 (2006)

\bibitem{ASH07}
P.~Ashwin, G.~Orosz, J.~Wordsworth, S.~Townley, SIAM J.~Appl. Dyn. Syst.
  \textbf{6}, 728 (2007)

\bibitem{FLU10b}
V.~Flunkert, S.~Yanchuk, T.~Dahms, E.~Sch{\"o}ll, Phys.~Rev.~Lett.
  \textbf{105}, 254101 (2010)

\bibitem{LEH11}
J.~Lehnert, T.~Dahms, P.~H{\"o}vel, E.~Sch{\"o}ll, Europhys. Lett. \textbf{96},
  60013 (2011)

\bibitem{ENG11}
A.~Englert, S.~Heiligenthal, W.~Kinzel, I.~Kanter, Phys. Rev.~E \textbf{83},
  046222 (2011)

\bibitem{KAN11}
I.~Kanter, M.~Zigzag, A.~Englert, F.~Geissler, W.~Kinzel, Europhys.~Lett.
  \textbf{93}, 60003 (2011)

\bibitem{OME11}
I.~Omelchenko, Y.L. Maistrenko, P.~H{\"o}vel, E.~Sch{\"o}ll, Phys. Rev. Lett.
  \textbf{106}, 234102 (2011)

\bibitem{DAH12}
T.~Dahms, J.~Lehnert, E.~Sch{\"o}ll, Phys. Rev.~E \textbf{86}, 016202 (2012)

\bibitem{HAG12}
A.~Hagerstrom, T.E. Murphy, R.~Roy, P.~H{\"o}vel, I.~Omelchenko, E.~Sch{\"o}ll,
  Nature Physics \textbf{8}, 658 (2012), published online

\bibitem{NEP12}
T.~Nepusz, T.~Vicsek, Nature Physics \textbf{8}, 568 (2012)

\bibitem{PER68}
D.H. Perkel, T.H. Bullock, Neurosci Res Program Bull \textbf{6}, 221 (1968)

\bibitem{WAT98}
D.J. Watts, S.H. Strogatz, Nature \textbf{393}, 440 (1998)

\bibitem{ADA99}
L.A. Adamic, \emph{The Small World Web}, Vol. 1696/1999 of \emph{Lecture Notes
  in Computer Science} (Springer Berlin / Heidelberg, 1999), ISBN
  978-3-540-66558-8

\bibitem{SPO00}
O.~Sporns, G.~Tononi, G.M. Edelman, Cereb. Cortex \textbf{10}, 127 (2000)

\bibitem{SPO06}
O.~Sporns, Biosystems \textbf{85}, 55 (2006)

\bibitem{LAT01}
V.~Latora, M.~Marchiori, Phys. Rev. Lett. \textbf{87}, 198701 (2001)

\bibitem{TRA82}
R.D. Traub, R.K. Wong, Science \textbf{216}, 745 (1982)

\bibitem{ROE97}
P.~Roelfsema, A.~Engel, P.~König, W.~Singer, Nature \textbf{385}, 157 (1997)

\bibitem{ENG01a}
A.~Engel, P.~Fries, W.~Singer, Nature Reviews Neuroscience \textbf{2}, 704
  (2001)

\bibitem{PIK01}
A.S. Pikovsky, M.G. Rosenblum, J.~Kurths, \emph{Synchronization, A Universal
  Concept in Nonlinear Sciences} (Cambridge University Press, Cambridge, 2001)

\bibitem{MEL07}
L.~Melloni, C.~Molina, M.~Pena, D.~Torres, W.~Singer, E.~Rodriguez,
  J.~Neurosci. \textbf{27}, 2858 (2007)

\bibitem{HAI06}
B.~Haider, A.~Duque, A.R. Hasenstaub, D.A. McCormick, J. Neurosci. \textbf{26},
  4535 (2006)

\bibitem{NEW99b}
M.E.J. Newman, D.J. Watts, Phys. Lett. A \textbf{263}, 341 (1999)

\bibitem{IZH00}
E.M. Izhikevich, Int. J. Bifurcation Chaos \textbf{10}, 1171 (2000)

\bibitem{HOD48}
A.L. Hodgkin, J. Physiol. \textbf{107}, 165 (1948)

\bibitem{HU93a}
G.~Hu, T.~Ditzinger, C.Z. Ning, H.~Haken, Phys.~Rev.~Lett. \textbf{71}, 807
  (1993)

\bibitem{HIZ07}
J.~Hizanidis, R.~Aust, E.~Sch{\"o}ll, Int.~J.~Bifur.~Chaos \textbf{18}, 1759
  (2008)

\bibitem{AUS09}
R.~Aust, P.~H{\"o}vel, J.~Hizanidis, E.~Sch{\"o}ll, Eur. Phys.~J.~ST
  \textbf{187}, 77 (2010)

\bibitem{ERD59}
P.~Erd\H{o}s, A.~R\'{e}nyi, Publ. Math. Debrecen \textbf{6}, 290 (1959)

\bibitem{MON99}
R.~Monasson, Eur. Phys. J. B \textbf{12}, 555 (1999)

\bibitem{PEC98}
L.M. Pecora, T.L. Carroll, Phys. Rev. Lett. \textbf{80}, 2109 (1998)

\bibitem{GER31}
S.A. Gerschgorin, Izv. Akad. Nauk. SSSR \textbf{7}, 749 (1931)

\bibitem{FAR01}
I.~Farkas, I.~Derenyi, A.L. Barab\'asi, T.~Vicsek, Phys. Rev.~E \textbf{64},
  026704 (2001)

\bibitem{CHO09}
C.U. Choe, T.~Dahms, P.~H{\"o}vel, E.~Sch{\"o}ll, Phys. Rev.~E \textbf{81},
  025205(R) (2010)

\bibitem{CHO11}
C.U. Choe, T.~Dahms, P.~H{\"o}vel, E.~Sch{\"o}ll, \emph{Control of synchrony by
  delay coupling in complex networks}, in \emph{Proceedings of the Eighth AIMS
  International Conference on Dynamical Systems, Differential Equations and
  Applications} ({American Institute of Mathematical Sciences}, Springfield,
  MO, USA, 2011), pp. 292--301, {DCDS} Supplement Sept. 2011

\end{thebibliography}

\end{document}